\documentclass[11pt]{article}
\usepackage[utf8]{inputenc}

\newcommand{\blind}{0}

\usepackage{caption}

\usepackage[a4paper, total={6.5in, 9in}]{geometry}
\usepackage{amsmath,amssymb,amsthm,bm}
\usepackage{graphicx,psfrag,epsf}
\usepackage{multicol}
\usepackage{flafter}
\usepackage{float}
\usepackage{setspace}
\usepackage{longtable}
\usepackage{rotating}
\usepackage{pdfpages}

\usepackage{color,hyperref,xcolor}
\hypersetup{colorlinks=true,urlcolor=purple,citecolor=blue,linkcolor=blue}
\usepackage{cleveref}
\usepackage[ruled,vlined]{algorithm2e}
\usepackage{subcaption}
\usepackage{comment}
\usepackage{booktabs}
\usepackage{multirow}
\usepackage{listings}
\usepackage{xcolor}

\definecolor{codegreen}{rgb}{0,0.6,0}
\definecolor{codegray}{rgb}{0.5,0.5,0.5}
\definecolor{codepurple}{rgb}{0.58,0,0.82}
\definecolor{backcolour}{rgb}{0.95,0.95,0.92}

\lstdefinestyle{mystyle}{
    backgroundcolor=\color{white},   
    commentstyle=\color{black},
    keywordstyle=\color{black},
    numberstyle=\tiny\color{black},
    stringstyle=\color{black},
    basicstyle=\ttfamily\footnotesize,
    breakatwhitespace=false,         
    breaklines=true,                 
    captionpos=b,                    
    keepspaces=true,                 
    numbers=left,                    
    numbersep=5pt,                  
    showspaces=false,                
    showstringspaces=false,
    showtabs=false,                  
    tabsize=2
}

\usepackage{enumerate}
\usepackage{url} 
\allowdisplaybreaks

\usepackage{booktabs}
\usepackage{graphicx}
\usepackage{ulem}

\numberwithin{equation}{section}

\newtheoremstyle{general}
{3mm} 
{3mm} 
{} 
{} 
{\bfseries} 
{.} 
{.5em} 
{} 

\theoremstyle{general}

\theoremstyle{definition}

\theoremstyle{remark}



\date{}

\begin{document}

\allowdisplaybreaks

\def\spacingset#1{\renewcommand{\baselinestretch}%
{#1}\small\normalsize} \spacingset{1}



*Subject Section*

# MR.RGM: An R Package for Fitting Bayesian Multivariate Bidirectional Mendelian Randomization Networks

Bitan Sarkar[1], Yang Ni[1,*]

[1]Department of Statistics, Texas A&M University, 3143, College Station, TX 77843, USA,

*To whom correspondence should be addressed.

## Abstract

**Motivation:** Mendelian randomization (MR) infers causal relationships between exposures and outcomes using genetic variants as instrumental variables. Typically, MR considers only a pair of exposure and outcome at a time, limiting its capability of capturing the entire causal network. We overcome this limitation by developing '**MR.RGM**' [1] (Mendelian randomization via reciprocal graphical model), a fast R-package that implements the Bayesian reciprocal graphical model [2] and enables practitioners to construct holistic causal networks with possibly cyclic/reciprocal causation and proper uncertainty quantifications, offering a comprehensive understanding of complex biological systems and their interconnections.

**Results:** We developed '**MR.RGM**', an open-source R package that applies bidirectional MR using a network-based strategy, enabling the exploration of causal relationships among multiple variables in complex biological systems. '**MR.RGM**' holds the promise of unveiling intricate interactions and advancing our understanding of genetic networks, disease risks, and phenotypic complexities.

**Availability:** '**MR.RGM**' is available at CRAN and https://github.com/bitansa/MR.RGM.

**Contact:** yni@stat.tamu.edu

**Supplementary information:** Supplementary data are available at *Bioinformatics* online.

## 1 Introduction

Mendelian randomization (MR) is a vital tool in genetic and epidemiological studies, emulating randomized control trials to elucidate causal relationships between exposures and outcomes. MR utilizes genetic variants as instrumental variables to mitigate confounding biases in observational studies. While traditional MR focuses on individual exposures and specific outcomes, the challenge arises in exploring multifaceted causal relationships involving multiple outcomes in complex biological systems.

A comprehensive review of existing R packages for MR reveals a range of tools with distinct functionalities. The package '**mr.pivw**' [3] implements the penalized inverse-variance weighted (pIVW) estimator, designed to estimate causal effects using summary-level data from genome-wide association studies. '**mr.raps**' [4] utilizes the Robust Adjusted Profile Score (RAPS) method for two-sample MR, enhancing causal inference through summary statistics. '**PPMR**' [5] excels in efficient two-sample MR analysis, addressing issues related to correlated instruments and horizontal pleiotropy.

'**OneSampleMR**' [6] is tailored for one-sample MR and instrumental variable analyses, leveraging individual-level data for detailed analyses. '**MRPC**' [7] applies principal component analysis (PCA) within the MR framework, facilitating multivariate MR by summarizing complex relationships into principal components. '**MendelianRandomization**' [8] supports both univariate and multivariate approaches using summary data, providing a versatile tool for complex MR analyses.

Recent advancements include '**TwoSampleMR**' [9], designed for robust two-sample MR analyses, and '**MVMR**' [10], which efficiently handles multivariate MR by modeling multiple outcomes and exposures. '**mrbayes**' [11] provides Bayesian implementation of multivariate MR based on either the inverse variance weighted model [12] or the Egger regression model [13], enabling uncertainty quantification through

MCMC sampling. **'MrDAG'** [14, 15] uses directed acyclic graphs to elucidate causal structures, focusing on causal inference within a Bayesian network framework.

Each package contributes significantly to MR research with distinct functionalities tailored to different data scenarios. However, common limitations include challenges in uncertainty quantification, restrictions to specific data types, constraints to bivariate analyses, and limitation in application to bidirectional MR. For details, refer to Table 1.

| R Package | Multivariate MR | Summary Level Data | Uncertainty Quantification | Bidirectionality |
|---|---|---|---|---|
| **mr.pivw** | ✗ | ✓ | ✓ | ✗ |
| **mr.raps** | ✓ | ✓ | ✓ | ✗ |
| **PPMR** | ✓ | ✓ | ✓ | ✗ |
| **MRPC** | ✓ | ✓ | ✓ | ✗ |
| **OneSampleMR** | ✗ | ✗ | ✓ | ✗ |
| **Mendelian Randomization** | ✓ | ✓ | ✓ | ✗ |
| **TwoSampleMR** | ✓ | ✓ | ✓ | ✗ |
| **MVMR** | ✓ | ✓ | ✓ | ✗ |
| **mrbayes** | ✓ | ✓ | ✓ | ✗ |
| **MrDAG** | ✓ | ✓ | ✓ | ✗ |
| **MR.RGM** | ✓ | ✓ | ✓ | ✓ |

Table1. "✓" indicates that a package is capable of performing a specific Sutask, while "✗" denotes that the package is not equipped to handle the task. Note that even if a package is not designed specifcially for bidirectional MR (indicated by "✗"), it can be in principle repeatedly applied to infer bi-directional causation.

This work introduces **'MR.RGM'** (for modeling details, refer to Supplementary Materials, Section 1), a new package for multivariate bidirectional MR based on reciprocal graphical model. '**MR.RGM**' employs a Bayesian framework and a network-based strategy, surpassing traditional MR approaches. While conventional methods focus on isolated variables, '**MR.RGM**' constructs comprehensive causal networks possibly with feedback loops, particularly adept at capturing intricate relationships within complex biological systems. '**MR.RGM**' supports individual-level data and two types of summary-level data, leveraging C++ for backend computations and the Woodbury matrix identity for efficiency (refer to Supplementary Materials, Section 6 & 7). Its Bayesian nature enables natural uncertainty quantification.

## 2 Methods

The R package '**MR.RGM**' has two key functions: **RGM**, dedicated to building causal graphs and estimating causal effects, and **NetworkMotif**, designed for uncertainty quantification, providing posterior probabilities for specific network motifs. First, we delve into the functionality of the **RGM** function, followed by a discussion of the **NetworkMotif** function. For detailed implementation and usage examples, see Section 5 of the Supplementary Materials.

**Function 1: RGM**

**Three Input Formats:**
**RGM** supports three input formats, catering to various data availability scenarios.

**Individual-Level Data Input:**
Researchers input *X* and *Y* matrices. *X* signifies the instrument data matrix, where columns represent instruments (e.g., SNPs), and rows represent observations. *Y* denotes the response data matrix, with columns representing response variables (e.g., protein, gene, or phenotype), and rows corresponding to observations.

**Summary-Level Data Input:**
Recognizing that individual-level data, especially SNPs or disease status, can be sensitive and not always readily available, RGM provides support for two types of summary-level data input. The first type involves specifying *Syy*, *Syx*, and *Sxx* matrices. *Syy* represents the covariance matrix of the response variables, *Syx* signifies the covariance matrix between the responses and instruments, and *Sxx* denotes the covariance matrix of the instrument variables. **RGM** would output the exact same solution as that based on individual-level data. For the calculation of the likelihood using summary-level data, refer to Supplementary Materials, Section 9.

**Handling Difficult Cross-Correlations:**
In situations where obtaining cross-correlations among response variables (*Syy*) is challenging, for instance, when two response variables were never measured in the same dataset, RGM can still be used and will output an approximate solution. In this scenario, researchers can specify *Sxx*, *Beta*, and *SigmaHat* matrices as input. *Beta* and *SigmaHat* matrices respectively contain regression coefficients and mean square errors for the regression of a response variable on its corresponding instrument, where the regression is performed without including an intercept term. Each row of *Beta* and *SigmaHat* matrices corresponds to a response variable, and each column corresponds to an instrument. For the detailed derivation, please refer to Section 10 and Section 11 of the Supplementary Materials.

**Essential Input Parameters:**
Users need to specify two input parameters: *D* and *n*. The matrix *D* is binary with dimensions p x k, where p is the number of response variables and k is the number of instrumental variables. Each row represents a response variable, and each column represents an instrumental variable. A 1 in *D*[i, j] indicates that instrumental variable j affects response variable i. For identifiability (refer to Supplementary Materials, Section 4), each response must be influenced by at least one instrumental variable that affects only that response, meaning there must be at least one unique entry of 1 in each row, where the corresponding column contains 0 in all other rows. If this condition is violated, the algorithm will throw an error when using *Syy*, *Beta* and *SigmHat*, but will issue a warning and proceed when using *X*, *Y*, or *Syy*, *Syx*, *Sxx*. The parameter *n* is the sample size (only needed for summary-level data).

**Customizable Parameters:**
RGM offers customizable analysis through user-defined parameters. Users can specify the number of MCMC sampling iterations (*nIter*) with a default of 10,000, set the number of discarded samples (*nBurnin*) with a default of 2,000, and define the thinning of posterior samples using the *Thin* parameter (default value: 1). Additionally, users can choose between two graph structure priors: "*Spike and Slab*" or "*Threshold*," with the default being "*Spike and Slab*". For details of these priors, refer to Supplementary Materials, Section 2.

Users also have the option to customize the model hyperparameters *aRho*, *bRho*, *nu1*, *aPsi*, *bPsi*, *nu2*, *aSigma*, *bSigma*, *PropVarA*,

*PropVarB*), although relatively robust default values are provided for ease of use. For details, see Section 3 of the Supplementary Materials.

**Outputs:**

The main outputs of **RGM** encompass causal effect estimates (*AEst* and *BEst*) among response variables and between instruments and responses. Additionally, **RGM** outputs binary adjacency matrices (*zAEst* and *zBEst*) that depict graph structures, derived by thresholding the posterior probability matrices (*GammaEst* and *PhiEst* respectively). The thresholding process may result in some entries of *zAEst* being zero even when the corresponding entries in *AEst* are non-zero. This discrepancy arises because *AEst* represents the posterior mean of causal effects, while *zAEst* reflects whether the inclusion probability for an edge exceeds the threshold. For practical use, multiplying *AEst* by *zAEst* elementwise can help account for edge inclusion probabilities, effectively making causal effects zero where inclusion probabilities are very small. RGM further provides posterior samples of the adjacency matrix between response variables (*GammaPst*), an input for the **NetworkMotif** function to quantify uncertainty for a specific network motif. Other outputs comprise of *A0Est*, *B0Est*, *GammaEst*, *TauEst*, *RhoEst*, *PhiEst*, *EtaEst*, *PsiEst*, *tAEst*, *tBEst*, *SigmaEst*, *AccptA*, *AccptB*, *AccpttA*, *AccpttB* and *LLPst*. All these outputs represent posterior means except *LLPst*, which consists of posterior samples; see Section 3 of the Supplementary Materials for details. Next, we will describe the **NetworkMotif** function.

**Function 2: NetworkMotif**

**Inputs:**

The **NetworkMotif** function requires two input variables: *Gamma* and *GammaPst*. *Gamma* represents a specific adjacency matrix corresponding to the network motif of the response variables for which uncertainty quantification is desired. *GammaPst* consists of the posterior samples of the adjacency matrix obtained from the **RGM** function.

**Output:**

The output of the **NetworkMotif** function includes the uncertainty quantification (i.e., posterior probability) of the specified network motif. For details, please refer to Section 3 of the Supplementary Materials.

## 3 Simulation Results

Utilizing the mathematical model y = Ay + Bx + e, we conducted a comprehensive assessment of our algorithm's performance. By generating x and e from a normal distribution, we simulated various scenarios to gauge the effectiveness of our approach. We compared our package with **'OneSampleMR'**, a recent and advanced R package for MR, based on true positive rate (TPR), false positive rate (FPR), false discovery rate (FDR), Matthew's correlation coefficient (MCC), and area under the receiver operating characteristic curve (AUC) at different sample sizes (10k, 30k, 50k), network sizes (5, 10), sparsity of A (25%, 50%), and effect sizes (0.1, -0.1). We adjust the standard deviation of the noise to achieve different variance explained levels (1%, 3%, 5%, and 10%). B is set to be the identity matrix. The detailed results are presented in the Supplementary Materials, Section 12 & Table 2, 3, 4, 5. Notably, our approach consistently yielded equivalent results when applied to both individual-level data and summary-level data, validating its versatility and reliability across different data availability scenarios. As

an example of comparison (Figure 1), it is evident that the networks produced by our method, exhibit a substantial improvement in accuracy compared to those generated by the package **'OneSampleMR'**. For further details on the selection of instrumental variables in real-world data and the application of these methods to GTEx [16] V7 data, please refer to Supplementary Materials, Section 13. In addition, we evaluated **MR.RGM**'s sensitivity to the assumption of normal errors by comparing performance across various error distributions, including normal, t-distributions, and Laplace. The results, indicating that '**MR.RGM**' is not sensitive to the normal error assumption, are detailed in Supplementary Materials, Section 8 & Table 1.

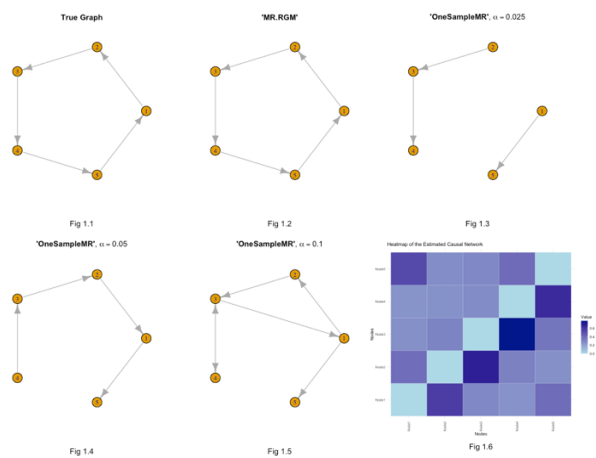

**Fig. 1.** Fig 1.1 represents the true graph. Fig 1.2 illustrates the causal network estimated by **'MR.RGM'** with the Spike and Slab prior. Fig 1.3, 1.4, and 1.5 depict the causal graph estimated by **'OneSampleMR'** with different alpha values. Fig 1.6 displays the heatmap of the causal graph created by **'MR.RGM'**. In the heatmap, each entry corresponds to node i in the horizontal row and node j in the vertical row, indicating the posterior probability of the causal effect of node i on node j. Lighter colors signify lower probabilities of any causal effect.

## 4 Conclusion

This paper introduces '**MR.RGM**', an R package for exploring causal relationships in complex biological systems. '**MR.RGM**' constructs comprehensive causal graphs possibly with feedback loops, accommodating both individual and summary-level data. Its versatility, adaptability to diverse data formats, and network-based strategy enhance causal inference. **MR.RGM**'s Bayesian approach enables uncertainty quantification and holds the promise to advance genetics and epidemiology research by uncovering complex biological phenomena.

## Acknowledgements



\if0\blind
{
  \title{\bf SUPPLEMENTARY MATERIAL\\
  MR.RGM: An R Package for Fitting Bayesian Multivariate Bidirectional Mendelian Randomization Networks}
  \author{$\text{Bitan Sarkar}\textsuperscript{1}$ and $\text{Yang Ni}\textsuperscript{1}$\\
     \\
    \textsuperscript{1} Department of Statistics, Texas A$\&$M University, 3143, College Station, TX 77843, USA}
  \maketitle
} \fi

\section{Model}
\label{sec:Model}

Let $\mathbf{Y_i} = (Y_{i1}, \cdots, Y_{ip})^{T}$ denotes the expressions for response variables $1, \cdots, p$, let $\mathbf{X_i} = (X_{i1}, \cdots, X_{ik})^{T}$ be the set of measurements for instrument variables $1, \cdots, k$ and for $i = 1, \ldots, n$. We have the following model:
\begin{align} \label{Eq1}
    & \mathbf{Y_i} = \mathbf{AY_i} + \mathbf{BX_i} + \mathbf{E_i}, \quad \mathbf{E_i} \sim N(0, \mathbf{\Sigma}) \nonumber \\
    \implies & (\mathbf{I_p} - \mathbf{A}) \mathbf{Y_i} = \mathbf{BX_i} + \mathbf{E_i} \nonumber \\
    \implies & \mathbf{Y_i} = (\mathbf{I_p} - \mathbf{A})^{-1}\mathbf{BX_i} + (\mathbf{I_p} - \mathbf{A})^{-1} \mathbf{E_i}
\end{align}
where $\mathbf{A} = (a_{ij}) \in \mathbf{R}^{p \times p}$ with zeroes on the diagonal, denotes the causal effects or strengths between the response variables, $\mathbf{B} = (b_{il}) \in \mathbf{R}^{p \times k}$ denotes the causal effects or strengths of the instrument variables on the response variables, $\mathbf{E_i} = (\epsilon_{1i}, \cdots, \epsilon_{pi})^{T} \sim N_p(0, \mathbf{\Sigma})$ and $\mathbf{E_i}$ and $\mathbf{X_i}$ are independent for $i = 1, \ldots, n$. $\mathbf{I_p}$ denotes a $p \times p$ identity matrix. We assume that $\mathbf{\Sigma}$ is diagonal i.e. $\mathbf{\Sigma} = \text{diag}(\sigma_1, \cdots, \sigma_p)$, i.e. the residuals are independent. Thus model (\ref{Eq1}) can be equivalently expressed as:
\begin{align}
    \mathbf{Y_i}|\mathbf{X_i} \sim N_p \{(\mathbf{I}_p - \mathbf{A})^{-1}\mathbf{BX_i}, (\mathbf{I}_p - \mathbf{A})^{-1}\mathbf{\Sigma}(\mathbf{I}_p - \mathbf{A})^{-T}\}
\end{align}

\section{Prior Specification}
\label{sec:Prior}

\textbf{MR.RGM} \cite{CRAN, ni2018reciprocal} offers the capability to estimate graph structures both among response variables and between response and instrument variables. This estimation process accommodates two distinct prior assumptions: ``\textbf{Threshold}" and ``\textbf{Spike and Slab}" priors, which users can specify through the ``\textbf{prior}'' parameter in \textbf{RGM}. The prior assumptions significantly influence the parameter settings and the full conditionals employed during the MCMC sampling procedure. We have the following model:
\begin{align} \label{Eq2}
    & \mathbf{Y_i} = \mathbf{AY_i} + \mathbf{BX_i} + \mathbf{E_i}, \,\, i =1, \ldots, n,
\end{align}
where $\mathbf{A} = (a_{ij}) \in \mathbf{R}^{p \times p}$ with zeroes on the diagonal, denotes the causal effect or strength between the response variable, $\mathbf{B} = (b_{il}) \in \mathbf{R}^{p \times k}$ denotes the causal effects or strengths of instrument variables on response variables, $\mathbf{E_i} = (\epsilon_{1i}, \cdots, \epsilon_{pi})^{T} \sim N_p(0, \mathbf{\Sigma})$ and $\mathbf{E_i}$ and $\mathbf{X_i}$ are independent, for $i=1,\ldots,n$. We assume that $\mathbf{\Sigma}$ is diagonal, i.e. $\mathbf{\Sigma} = \text{diag}(\sigma_1, \cdots, \sigma_p)$, i.e. the residuals are independent. Here's an overview of the prior assumptions on the model parameters and their impact on the analysis:

\subsection{Threshold Prior}
\label{sec:Threshold Prior}

We have the following prior assumptions in this case:
\begin{align} \label{Eq 3.2}
\begin{split}
    \Tilde{a}_{ij} & \sim N(0, \tau_{ij}), \quad (i = 1, \ldots, p;\, j =1, \ldots, p)\\
    a_{ij} & = \Tilde{a}_{ij}\mathbf{I}(|\Tilde{a}_{ij}| > t_{A}), \quad (i = 1, \ldots, p;\, j =1, \ldots, p)\\
    t_A & \sim Uniform(0, 1)\\
    \sqrt{\tau_{ij}} & \sim C^{+}(0, 1),\quad (i = 1, \ldots, p;\, j =1, \ldots, p)\\\\
    \Tilde{b}_{il} & \sim N(0, \eta_{il}), \quad (i = 1, \ldots, p;\, l =1, \ldots, k)\\
    b_{il} & = \Tilde{b}_{il}\mathbf{I}(|\Tilde{b}_{il}| > t_{B}), \quad (i = 1, \ldots, p;\, l =1, \ldots, k)\\
    t_B & \sim Uniform(0, 1)\\
    \sqrt{\eta_{il}} & \sim C^{+}(0, 1),\quad (i = 1, \ldots, p;\, l =1, \ldots, k)\\\\
    \sigma_i & \sim IG(a_{\sigma}, b_{\sigma}),\quad (i = 1, \ldots, p)
\end{split}
\end{align}
where $C^{+}(0, 1)$ denotes standard half-cauchy distribution and $x \sim C^{+}(0, 1)$ implies the following:
\begin{align*}
    x^2|a \sim IG(1/2, 1/a) \quad \text{and} \quad a \sim IG(1/2, 1)
\end{align*}
Posterior inference under Threshold prior is implemented by the following MCMC posterior simulation.
\begin{enumerate}
    \item Update $\eta_{il}$ by a Gibbs transition probability. Draw $\epsilon \sim IG(1, 1 + 1 / \eta_{il})$ and then draw $\eta_{il} \sim IG(1, \tilde{b}_{il}^2 / 2 + 1 / \epsilon)$.
   
    \item Update $\Tilde{b}_{il}$ and $b_{il}$ by a random walk  Metropolis-Hastings (M-H) transition probability. Draw $\Tilde{b}_{il}^{*} \sim N(\Tilde{b}_{il}, \textbf{Prop{VarB}})$ and create $\mathbf{B^{*}}$ from $\mathbf{B}$ by substituting $b_{il}$ by $b_{il}^{*}$ where $b_{il}^{*} = \Tilde{b}_{il}^{*}\mathbf{I}(|\Tilde{b}_{il}^{*}| > t_{B})$. Accept $\Tilde{b}_{il}^{*}$ and $b_{il}^{*}$ with probability min$(\alpha, 1)$ where,
    \begin{align*}
        \alpha = \frac{p(\mathbf{Y|X, A, B^{*}, \Sigma})p(\Tilde{b}_{il}^{*}|\eta_{il})}{p(\mathbf{Y|X, A, B, \Sigma})p(\Tilde{b}_{il}|\eta_{il})}.
    \end{align*}

    \item Update $\phi_{il}$ based on the value of $b_{il}$. Make $\phi_{il} = 1$ if $b_{il} \neq 0$, else, make $\phi_{il} = 0$.

    \item  Update $t_B$ by an M-H transition probability with a truncated normal proposal. Draw $t_B^{*} \sim TN(t_B, 0.01, 0, 1)$ where $TN(t_B, 0.01, 0, 1)$ denotes a normal distribution with mean $t_B$, variance $0.01$ and restricted to the interval $(0, 1)$ and accept it with probability min$(\alpha, 1)$ where,
     \begin{align*}
        \alpha = \frac{p(\mathbf{Y|X, A, B^{*}, \Sigma})TN(t_B|t_B^{*}, 0.01, 0, 1)}{p(\mathbf{Y|X, A, B, \Sigma})TN(t_B^{*}|t_B, 0.01, 0, 1)}.
    \end{align*}
    $\mathbf{B^{*}}$ is obtained by thresholding $\mathbf{\Tilde{B}}$ at $t_B^{*}$.

    \item Update $\sigma_{j}$ by a Gibbs transition probability. Draw $\sigma_j \sim IG(a_{\sigma} + n / 2, b_{\sigma} + SSE_{j} / 2)$ where,
    \begin{align*}
        SSE_{j} = & n \times (\mathbf{I_p - A)_{(j,.)}} \times \mathbf{Syy} \times (\mathbf{I_p - A)_{(j,.)}}^{T} - 2 \times n \times (\mathbf{I_p - A)_{(j,.)}} \times \mathbf{Syx} \times \mathbf{B_{(j,.)}}^{T}\\
        & + n \times \mathbf{B_{(j,.)}} \times \mathbf{Sxx} \times \mathbf{B_{(j,.)}}^{T}.
    \end{align*}
    $\mathbf{I_p}$ denotes a $p \times p$ identity matrix, $(\mathbf{I_p - A)_{(j,.)}}$ denotes $j$th row of $(\mathbf{I_p - A})$ and $\mathbf{B_{(j,.)}}$ denotes $j$th row of $\mathbf{B}$.

    \item Update $\tau_{ij}$ by a Gibbs transition probability. Draw $\epsilon \sim IG(1, 1 + 1 / \tau_{ij})$ and then draw $\tau_{ij} \sim IG(1, \tilde{a}_{ij}^2 / 2 + 1 / \epsilon)$.
   
    \item Update $\Tilde{a}_{ij}$ and $a_{ij}$ by a random walk  Metropolis-Hastings (M-H) transition probability. Draw $\Tilde{a}_{ij}^{*} \sim N(\Tilde{a}_{ij}, \textbf{Prop{VarA}})$ and create $A^{*}$ from $A$ by substituting $a_{ij}$ by $a_{ij}^{*}$ where $a_{ij}^{*} = \Tilde{a}_{ij}^{*}\mathbf{I}(|\Tilde{a}_{ij}^{*}| > t_{A})$. Accept $\Tilde{a}_{ij}^{*}$ and $a_{ij}^{*}$ with probability min$(\alpha, 1)$ where,
    \begin{align*}
        \alpha = \frac{p(\mathbf{Y|X, A^{*}, B, \Sigma})p(\Tilde{a}_{ij}^{*}|\tau_{ij})}{p(\mathbf{Y|X, A, B, \Sigma})p(\Tilde{a}_{ij}|\tau_{ij})}.
    \end{align*}

    \item Update $\gamma_{ij}$ based on the value of $a_{ij}$. Make $\gamma_{ij} = 1$ if $a_{ij} \neq 0$, else, make $\gamma_{ij} = 0$.

    \item  Update $t_A$ by an M-H transition probability with a truncated normal proposal. Draw $t_A^{*} \sim TN(t_A, 0.01, 0, 1)$ where $TN(t_A, 0.01, 0, 1)$ denotes a normal distribution with mean $t_A$, variance $0.01$ and restricted to the interval $(0, 1)$ and accept it with probability min$(\alpha, 1)$ where,
     \begin{align*}
        \alpha = \frac{p(\mathbf{Y|X, A^{*}, B, \Sigma})TN(t_A|t_A^{*}, 0.01, 0, 1)}{p(\mathbf{Y|X, A, B, \Sigma})TN(t_A^{*}|t_A, 0.01, 0, 1)}.
    \end{align*}
    $\mathbf{A^{*}}$ is obtained by thresholding $\mathbf{\Tilde{A}}$ at $t_A^{*}$.
\end{enumerate}

\subsection{Spike and Slab Prior}
\label{sec:Spike and Slab Prior}

We have the following prior assumptions in this case:
\begin{align} \label{Eq 3.3}
\begin{split}
 a_{ij} & \sim \gamma_{ij} N(0, \tau_{ij}) + (1 - \gamma_{ij}) N(0, \nu_1 \times \tau_{ij}),\quad (i = 1, \ldots, p;\, j =1, \ldots, p)\\
    \gamma_{ij} & \sim Ber(\rho_{ij}),\quad (i = 1, \ldots, p;\, j =1, \ldots, p)\\
    \rho_{ij} & \sim Beta(a_{\rho}, b_{\rho}),\quad (i = 1, \ldots, p;\, j =1, \ldots, p)\\
    \sqrt{\tau_{ij}} & \sim C^{+}(0, 1),\quad (i = 1, \ldots, p;\, j =1, \ldots, p)\\\\
    b_{il} & \sim \phi_{il} N(0, \eta_{il}) + (1 - \phi_{il}) N(0, \nu_2 \times \eta_{il}),\quad (i = 1, \ldots, p;\, l = 1, \ldots, k)\\
    \phi_{il} & \sim Ber(\psi_{il}),\quad (i = 1, \ldots, p;\, l = 1, \ldots, k)\\
    \psi_{il} & \sim Beta(a_{\psi}, b_{\psi}),\quad (i = 1, \ldots, p;\, l = 1, \ldots, k)\\
    \sqrt{\eta_{il}} & \sim C^{+}(0, 1),\quad (i = 1, \ldots, p;\, l =1, \ldots, k)\\\\
    \sigma_i & \sim IG(a_{\sigma}, b_{\sigma}),\quad (i = 1, \ldots, p)    
\end{split}
\end{align}
Posterior inference under Spike and Slab prior is implemented by the following MCMC posterior simulation.
\begin{enumerate}
    \item Update $\psi_{il}$ by a Gibbs transition probability. Draw $\psi_{il} \sim Beta(\phi_{il} + a_{\psi}, 1 - \phi_{il} + b_{\psi})$.
    
    \item Update $\eta_{il}$ by a Gibbs transition probability. Draw $\epsilon \sim IG(1, 1 + 1 / \eta_{il})$ and then draw $\eta_{il} \sim IG(1, b_{il}^2 / 2 + 1 / \epsilon)$ (if $\phi_{il} = 1$) or draw  $\eta_{il} \sim IG(1, b_{il}^2 / (2 \times \nu_2) + 1 / \epsilon)$ (if $\phi_{il} = 0$).
    
    \item Update $\phi_{il}$ by a Gibbs transition probability. Draw $\phi_{il} \sim Ber(p)$ where,
    \begin{align*}
        p = \frac{\exp{(- b_{il}^2 / (2 \times \eta_{il}))} \times \psi_{il}}{\exp{(- b_{il}^2 / (2 \times \eta_{il}))} \times \psi_{il} + \exp{(- b_{il}^2 / (2 \times \nu_2 \times \eta_{il}))} \times (1 - \psi_{il}) / \sqrt{\nu_2}}.
    \end{align*}
    
    \item Update $b_{il}$ by a random walk  Metropolis-Hastings (M-H) transition probability. Draw $b_{il}^{*} \sim N(b_{il}, \textbf{Prop{VarB}})$ and create $\mathbf{B^{*}}$ from $\mathbf{B}$ by substituting $b_{il}$ by $b_{il}^{*}$. Accept $b_{il}^{*}$ with probability min$(\alpha, 1)$ where,
    \begin{align*}
        \alpha = \frac{p(\mathbf{Y|X, A, B^{*}, \Sigma})p(b_{il}^{*}|\phi_{il}, \eta_{il}, \nu_2)}{p(\mathbf{Y|X, A, B, \Sigma})p(b_{il}|\phi_{il}, \eta_{il}, \nu_2)}.
    \end{align*}
    
    \item Update $\sigma_{j}$ by a Gibbs transition probability. Draw $\sigma_j \sim IG(a_{\sigma} + n / 2, b_{\sigma} + SSE_{j} / 2)$ where,
    \begin{align*}
        SSE_{j} = & n \times \mathbf{(I_p - A)_{(j,.)} \times Syy \times (I_p - A)_{(j,.)}^{T}} - 2 \times n \times \mathbf{(I_p - A)_{(j,.)} \times Syx \times B_{(j,.)}^{T}}\\
        & + n \times \mathbf{B_{(j,.)} \times Sxx \times B_{(j,.)}^{T}}.
    \end{align*}
    $\mathbf{I_p}$ denotes a $p \times p$ identity matrix, $\mathbf{(I_p - A)_{(j,.)}}$ denotes $j$th row of $\mathbf{(I_p - A)}$ and $\mathbf{B_{(j,.)}}$ denotes $j$th row of $\mathbf{B}$.
    
    \item Update $\rho_{ij}$ by a Gibbs transition probability. Draw $\rho_{ij} \sim Beta(\gamma_{ij} + a_{\rho}, 1 - \gamma_{ij} + b_{\rho})$.
    
    \item Update $\tau_{ij}$ by a Gibbs transition probability. Draw $\epsilon \sim IG(1, 1 + 1 / \tau_{ij})$ and then draw $\tau_{ij} \sim IG(1, a_{ij}^2 / 2 + 1 / \epsilon)$ (if $\gamma_{ij} = 1$) or draw  $\tau_{ij} \sim IG(1, a_{ij}^2 / (2 \times \nu_1) + 1 / \epsilon)$ (if $\gamma_{ij} = 0$).
    
    \item Update $\gamma_{ij}$ by a Gibbs transition probability. Draw $\gamma_{ij} \sim Ber(p)$ where
    \begin{align*}
        p = \frac{\exp{(- a_{ij}^2 / (2 \times \tau_{ij}))} \times \rho_{ij}}{\exp{(- a_{ij}^2 / (2 \times \tau_{ij}))} \times \rho_{ij} + \exp{(- a_{ij}^2 / (2 \times \nu_1 \times \tau_{ij}))} \times (1 - \rho_{ij}) / \sqrt{\nu_1}}.
    \end{align*}
    
    \item Update $a_{ij}$ by a random walk  Metropolis-Hastings (M-H) transition probability. Draw $a_{ij}^{*} \sim N(a_{ij}, \textbf{Prop{VarA}})$ and create $\mathbf{A^{*}}$ from $\mathbf{A}$ by substituting $a_{ij}$ by $a_{ij}^{*}$. Accept $a_{ij}^{*}$ with probability min$(\alpha, 1)$ where,
    \begin{align*}
        \alpha = \frac{p(\mathbf{Y|X, A^{*}, B, \Sigma})p(a_{ij}^{*}|\gamma_{ij}, \tau_{ij}, \nu_1)}{p(\mathbf{Y|X, A, B, \Sigma})p(a_{ij}|\gamma_{ij}, \tau_{ij}, \nu_1)}.
    \end{align*}
\end{enumerate}

\section{Available Functions}
\label{sec:Function}

\textbf{MR.RGM} provides two user-accessible functions: \textbf{RGM} and \textbf{NetworkMotif}. The primary objective of the \textbf{RGM} function is to establish causal networks among response variables and between response and instrument variables, while \textbf{NetworkMotif} focuses on quantifying uncertainty for a given network structure among the response variables. Detailed input and output structures for these functions, along with usage guidelines, are provided in the following subsections.

\subsection{RGM}
\subsubsection{Inputs}

We will now specify the inputs for \textbf{RGM} and give brief descriptions about them.
\begin{itemize}
    \item $\textbf{X}$: $\textbf{X}$ represents the instrument data matrix, which is $n \times k$, where each column corresponds to an instrument (e.g. DNA or SNPs), and each row represents a particular observation. The default value is set to \textbf{NULL}.
    \item $\textbf{Y}$: $\textbf{Y}$ represents the response data matrix, which is $n \times p$, where each column corresponds to a response (e.g. protein, gene or disease data), and each row represents a particular observation. The default value is set to \textbf{NULL}.
    \item $\textbf{Syy}$: $\textbf{Syy}$ represents the covariance matrix of the response variables. It is a matrix of dimensions $p \times p$. Here, ``$p$" signifies the number of response variables. This matrix is derived through the operation $(\textbf{Y}^{T} \times \textbf{Y}) / n$, where ``$\textbf{Y}$" denotes the response data matrix and ``$n$" stands for the total number of observations. 
    \item $\textbf{Syx}$: $\textbf{Syx}$ represents the covariance matrix between the responses and the instruments. It is a matrix of dimensions $p \times k$. Here, ``$p$" signifies the number of response variables and ``$k$" represents the number of instrument variables. This matrix is derived through the operation $(\textbf{Y}^{T} \times \textbf{X}) / n$, where ``$\textbf{Y}$" denotes the response data matrix, ``$\textbf{X}$" denotes the instrument data matrix and ``$n$" stands for the total number of observations. 
    \item $\textbf{Sxx}$: $\textbf{Sxx}$ represents the covariance matrix of the instrument variables. It is a matrix of dimensions $k \times k$. Here, ``$k$" signifies the number of instrument variables. This matrix is derived through the operation $(\textbf{X}^{T} \times \textbf{X}) / n$, where ``$\textbf{X}$" denotes the instrument data matrix and ``$n$" stands for the total number of observations. 
    \item $\textbf{Beta}$: $\textbf{Beta}$ is a matrix of dimensions $p \times k$. In this matrix, each row corresponds to a specific response variable, and each column pertains to a distinct instrument variable. Each entry within the matrix represents the regression coefficient of the individual response variable on the specific instrument variable, where the regression is performed without including an intercept term. 
     \item $\textbf{Sigma}\textbf{Hat}$: $\textbf{Sigma}\textbf{Hat}$ is a matrix of dimensions $p \times k$. In this matrix, each row corresponds to a specific response variable, and each column pertains to an individual instrument variable. Each entry in this matrix represents the mean square error associated with regressing the particular response on the specific instrument variable, where the regression is performed without including an intercept term. 
     \item $\textbf{D}$: $\textbf{D}$ is a binary indicator matrix of dimension $p \times k$, where $p$ represents the number of response variables, and $k$ represents the total number of instrumental variables. Each row of $\textbf{D}$ corresponds to a response variable, and each column corresponds to an instrumental variable. The entry $\textbf{D}[i,j]$ is set to $1$ if instrumental variable $j$ affects response variable $i$, and $0$ otherwise. For identifiability, each response variable must have at least one instrumental variable that affects only that specific response (refer to Section (\ref{Identifiable})). In other words, for each row in $\textbf{D}$, there must be at least one column with a $1$, and that column must contain $0$’s in all other rows. When using \textbf{Syy}, \textbf{Beta}, and \textbf{SigmaHat} as inputs, this condition must be strictly satisfied. If not, an error will be thrown, halting the algorithm. However, when using \textbf{X}, \textbf{Y}, or \textbf{Syy}, \textbf{Syx}, and \textbf{Sxx} as inputs, a warning will be issued if the condition is violated, but the method will proceed and users are cautioned to interpret the results.
     \color{black}
     \item $\textbf{n}$: $\textbf{n}$ is a positive integer input representing the number of datapoints or observations in the dataset. This input is only required when summary level data is given as input.
     \item $\textbf{nIter}$: $\textbf{nIter}$ is a positive integer input representing the number of MCMC (Markov Chain Monte Carlo) sampling iterations. The default value is set to $10,000$.
     \item $\textbf{nBurnin}$: $\textbf{nBurnin}$ is a non-negative integer input representing the number of samples to be discarded during the burn-in phase of MCMC sampling. It's important that $\textbf{nBurnin}$ is less than $\textbf{nIter}$. The default value is set to $2,000$.
     \item $\textbf{Thin}$: $\textbf{Thin}$ is a positive integer input denoting the thinning factor applied to posterior samples. Thinning reduces the number of samples retained from the MCMC process for efficiency. Thin should not exceed ($\textbf{nIter}$ - $\textbf{nBurnin}$). The default value is set to $1$.
     \item $\textbf{prior}$: $\textbf{prior}$ is a parameter representing the prior assumption on the graph structure. It offers two options: ``Threshold" or ``Spike and Slab". The default is set to ``Spike and Slab".
     \item $\textbf{aRho}$: $\textbf{aRho}$ is a positive scalar input representing the first parameter of a beta distribution in model (\ref{Eq 3.3}) for Spike and Slab prior assumption on $\textbf{A}$. The default value is set to $0.5$.
     \item $\textbf{bRho}$: $\textbf{bRho}$ is a positive scalar input representing the second parameter of a beta distribution in model (\ref{Eq 3.3}) for Spike and Slab prior assumption on $\textbf{A}$. The default value is set to $0.5$.
     \item $\textbf{nu1}$: $\textbf{nu1}$ is a positive scalar input representing the multiplication factor in the variance of the spike part in the spike and slab distribution of matrix $\textbf{A}$ in model (\ref{Eq 3.3}). The default value is set to $0.0001$.
     \item $\textbf{aPsi}$: $\textbf{aPsi}$ is a positive scalar input representing the first parameter of a beta distribution in model (\ref{Eq 3.3}) for Spike and Slab prior assumption on $\textbf{B}$. The default value is set to $0.5$.
     \item $\textbf{bPsi}$: $\textbf{bPsi}$ is a positive scalar input representing the second parameter of a beta distribution in model (\ref{Eq 3.3}) for Spike and Slab prior assumption on $\textbf{B}$. The default value is set to $0.5$.
     \item $\textbf{nu2}$: $\textbf{nu2}$ is a positive scalar input representing the multiplication factor in the variance of the spike part in the spike and slab distribution of matrix $\textbf{B}$ in model (\ref{Eq 3.3}). The default value is set to $0.0001$.
     \item $\textbf{aSigma}$: $\textbf{aSigma}$ is a positive scalar input corresponding to the first parameter of an inverse gamma distribution, which is associated to the variance of the model in model (\ref{Eq 3.2}) $\&$ (\ref{Eq 3.3}). The default value is set to $0.01$.
     \item $\textbf{bSigma}$: $\textbf{bSigma}$ is a positive scalar input corresponding to the second parameter of an inverse gamma distribution, which is associated to the variance of the model in model (\ref{Eq 3.2}) $\&$ (\ref{Eq 3.3}). The default value is set to $0.01$.
     \item $\textbf{Prop}\textbf{VarA}$: $\textbf{Prop}\textbf{VarA}$ is a positive scalar input corresponding to the variance of the normal distribution used for proposing terms within the $\textbf{A}$ matrix by Metropolis-Hastings (M-H) algorithm. The default value is set to $0.01$.
     \item $\textbf{Prop}\textbf{VarB}$: $\textbf{Prop}\textbf{VarB}$ is a positive scalar input corresponding to the variance of the normal distribution used for proposing terms within the $\textbf{B}$ matrix by Metropolis-Hastings (M-H) algorithm. The default value is set to $0.01$.
\end{itemize}

\textbf{Instructions to the Users:}
\begin{enumerate}
    \item \textbf{RGM} offers three data input options: individual-level data ($\textbf{X}$ and $\textbf{Y}$ matrices), summary-level data ($\textbf{Syy}$, $\textbf{Syx}$, $\textbf{Sxx}$) and a custom format ($\textbf{Sxx}$, $\textbf{Beta}$, $\textbf{Sigma}\textbf{Hat}$). It prioritizes $\textbf{X}$ and $\textbf{Y}$ matrices, then $\textbf{Syy}$, $\textbf{Syx}$, $\textbf{Sxx}$, and finally, $\textbf{Sxx}$, $\textbf{Beta}$, $\textbf{Sigma}\textbf{Hat}$. Users can select any of these data formats to suit their needs and don't have to specify all of them, allowing flexibility based on data availability. If none of these formats are provided, \textbf{RGM} prompts for valid data input.
    \item Users must provide their values for $\textbf{D}$ and $\textbf{n}$, two essential input parameters for \textbf{RGM}.
    \item Users have the flexibility to customize the rest input parameters in \textbf{RGM}, allowing them to fine-tune the model to better fit their data and research objectives. While default values are available, users can experiment with these parameters to assess their impact on model performance.
\end{enumerate}

\subsubsection{Outputs}

We will now specify outputs of \textbf{RGM} and will give brief descriptions about them.

\begin{itemize}
    \item {$\textbf{A}\textbf{Est}$}: {$\textbf{A}\textbf{Est}$} is a matrix of dimensions $p \times p$, representing the estimated causal effects or strengths between the response variables. Here ``$p$" signifies the number of response variable.
    \item {$\textbf{B}\textbf{Est}$}: {$\textbf{B}\textbf{Est}$} is a matrix of dimensions $p \times k$, representing the estimated causal effects or strengths between the response variables and the instrument variables. Each row corresponds to a specific response variable, and each column corresponds to a particular instrument variable. Here, ``$p$" signifies the number of response variables and ``$k$" represents the number of instrument variables.
    \item {$\textbf{zA}\textbf{Est}$}: {$\textbf{zA}\textbf{Est}$} is a binary adjacency matrix of dimensions $p \times p$, indicating the graph structure between the response variables. Each entry in the matrix represents the presence ($1$) or absence ($0$) of a causal link between the corresponding response variables. Here ``$p$" signifies the number of response variable.
    \item {$\textbf{zB}\textbf{Est}$}: {$\textbf{zB}\textbf{Est}$} is a binary adjacency matrix of dimensions $p \times k$, illustrating the graph structure between the response variables and the instrument variables. Here, ``$p$" signifies the number of response variables and ``$k$" represents the number of instrument variables. Each row corresponds to a specific response variable, and each column corresponds to a particular instrument variable. The presence of a causal link is denoted by $1$, while the absence is denoted by $0$.
    \item {$\textbf{A0}\textbf{Est}$}: {$\textbf{A0}\textbf{Est}$} is a matrix of dimensions $p \times p$, representing the estimated causal effects or strengths between response variables before thresholding. Here ``$p$" signifies the number of response variable. This output is particularly relevant for cases where the ``Threshold" prior assumption is utilized.
    \item {$\textbf{B0}\textbf{Est}$}: {$\textbf{B0}\textbf{Est}$} is a matrix of dimensions $p \times k$, representing the estimated causal effects or strengths between the response variables and the instrument variables before thresholding. This output is particularly relevant for cases where the ``Threshold" prior assumption is utilized. Here, ``$p$" signifies the number of response variables and ``$k$" represents the number of instrument variables. Each row corresponds to a specific response variable, and each column corresponds to a particular instrument variable.
    \item {\textbf{Gamma}\textbf{Est}}: {\textbf{Gamma}\textbf{Est}} is a matrix of dimensions $p \times p$, representing the estimated probabilities of edges between response variables in the graph structure. Here ``$p$" signifies the number of response variable. Each entry in the matrix indicates the probability of a causal link between the corresponding response variables. 
     \item {\textbf{Tau}\textbf{Est}}: {\textbf{Tau}\textbf{Est}} is a matrix of dimensions $p \times p$, representing the estimated variances of causal interactions between response variables. Here ``$p$" signifies the number of response variable. Each entry in the matrix corresponds to the variance of the causal effect between the corresponding response variables.  
    \item {\textbf{Phi}\textbf{Est}}: {\textbf{Phi}\textbf{Est}} is a matrix of dimensions $p \times k$, representing the estimated probabilities of edges between response and instrument variables in the graph structure. Here, ``$p$" signifies the number of response variables and ``$k$" represents the number of instrument variables. Each row corresponds to a specific response variable, and each column corresponds to a particular instrument variable. 
    \item {\textbf{Eta}\textbf{Est}}: {\textbf{Eta}\textbf{Est}} is a matrix of dimensions $p \times k$, representing the estimated variances of causal interactions between response and instrument variables. Here, ``$p$" signifies the number of response variables and ``$k$" represents the number of instrument variables. Each row corresponds to a specific response variable, and each column corresponds to a particular instrument variable.
    \item {\textbf{tA}\textbf{Est}}: {\textbf{tA}\textbf{Est}} is a scalar value representing the estimated thresholding value of causal interactions between response variables. This output is relevant when using the ``Threshold" prior assumption.
    \item {\textbf{tB}\textbf{Est}}: {\textbf{tB}\textbf{Est}} is a scalar value representing the estimated thresholding value of causal interactions between response and instrument variables. This output is applicable when using the ``Threshold" prior assumption.
    \item {\textbf{Sigma}\textbf{Est}}: {\textbf{Sigma}\textbf{Est}} is a vector of length $p$, representing the estimated variances of each response variable. Here ``$p$" signifies the number of response variables. Each element in the vector corresponds to the variance of a specific response variable. 
    \item \textbf{AccptA}: \textbf{AccptA} is the percentage of accepted entries in the $\textbf{A}$ matrix, which represents the causal interactions between response variables. This metric indicates the proportion of proposed changes that were accepted during the sampling process.
    \item \textbf{AccptB}: \textbf{AccptB} is the percentage of accepted entries in the $\textbf{B}$ matrix, which represents the causal interactions between response and instrument variables. This metric indicates the proportion of proposed changes that were accepted during the sampling process.
    \item {\textbf{Accpt}\textbf{tA}}: {\textbf{Accpt}\textbf{tA}} is the percentage of accepted thresholding values for causal interactions between response variables when using the ``Threshold" prior assumption. This metric indicates the proportion of proposed thresholding values that were accepted during the sampling process.
    \item {\textbf{Accpt}\textbf{tB}}: {\textbf{Accpt}\textbf{tB}} is the percentage of accepted thresholding values for causal interactions between response and instrument variables when using the ``Threshold" prior assumption. This metric indicates the proportion of proposed thresholding values that were accepted during the sampling process.
    \item {\textbf{LL}\textbf{Pst}}: {\textbf{LL}\textbf{Pst}} is a vector containing the posterior log-likelihoods of the model. Each element in the vector represents the log-likelihood of the model given the observed data and the estimated parameters.
   \item {\textbf{Gamma}\textbf{Pst}}:
   {\textbf{Gamma}\textbf{Pst}} comprises the posterior samples of the causal network among the response variables. It has dimensions $p\,\, \times p\,\, \times n_{pst}$, where $n_{pst}$ is the number of posterior samples and $p$ denotes the number of response variables. Each slice of the array represents a $p \times p$ matrix indicating the estimated causal network among the response variables at a particular iteration. This is a vital input for the \textbf{NetworkMotif} function to quantify uncertainty for a specific network.  
    
     \item {\textbf{Rho}\textbf{Est}}: {\textbf{Rho}\textbf{Est}} is a matrix of dimensions $p \times p$, representing the estimated Bernoulli success probabilities of causal interactions between response variables when using the ``Spike and Slab" prior assumption. Here ``$p$" signifies the number of response variable. Each entry in the matrix corresponds to the success probability of a causal interaction between the corresponding response variables.
    \item {\textbf{Psi}\textbf{Est}}: {\textbf{Psi}\textbf{Est}} is a matrix of dimensions $p \times k$, representing the estimated Bernoulli success probabilities of causal interactions between response and instrument variables when using the ``Spike and Slab" prior assumption. Here, ``$p$" signifies the number of response variables and ``$k$" represents the number of instrument variables. Each row in the matrix corresponds to a specific response variable, and each column corresponds to a particular instrument variable.
\end{itemize}

\subsection{NetworkMotif}

\subsubsection{Inputs}

Next, we'll outline the inputs required for \textbf{NetworkMotif} and provide concise descriptions for each of them.

\begin{itemize}
    \item  {\textbf{Gamma}}: \textbf{Gamma} represents a $p \times p$ matrix that signifies a specific network motif (i.e., the adjacency matrix of a subgraph) among the response variables, where ``$p$'' represents the number of response variables. This matrix is the focus of uncertainty quantification.
    \item {\textbf{Gamma}\textbf{Pst}}:
   {\textbf{Gamma}\textbf{Pst}} comprises the posterior samples of the causal network among the response variables. It has dimensions $p\,\, \times p\,\, \times n_{pst}$, where $n_{pst}$ is the number of posterior samples and $p$ denotes the number of response variables. Each slice of the array represents a $p \times p$ adjacency matrix indicating the estimated causal network among the response variables at a particular iteration. {\textbf{Gamma}\textbf{Pst}} may be obtained from the \textbf{RGM} function. 
\end{itemize}

\subsubsection{Output}

The \textbf{NetworkMotif} function calculates the posterior probability for the provided network motif as a measure of uncertainty. A value close to 1 indicates that the given network motif is frequently observed in the posterior samples, while a value close to 0 suggests that the given network structure is rarely observed in the posterior samples.


\section{Instrumental Variables in Mendelian Randomization} \label{Identifiable}

In Mendelian Randomization (MR) studies, the validity of instrumental variables (IVs) is fundamental for ensuring accurate causal inference. For an instrument to be considered valid, it must satisfy the following three key assumptions:

\begin{enumerate}
    \item \textbf{Relevance}: The instrument, denoted as $\mathbf{Z}$, must be associated with the exposure variable $\mathbf{X}$. This condition ensures that $\mathbf{Z}$ has a significant effect on $\mathbf{X}$, mathematically expressed as:
    \[
    \text{Cov}(\mathbf{Z}, \mathbf{X}) \neq 0.
    \]

    \item \textbf{Independence}: The instrument $\mathbf{Z}$ must be independent of any confounding variables $\mathbf{U}$ that influence both the exposure $\mathbf{X}$ and the outcome $\mathbf{Y}$. This ensures that $\mathbf{Z}$ is not related to any confounders that could introduce bias, which is expressed as:
    \[
    \text{Cov}(\mathbf{Z}, \mathbf{U}) = 0.
    \]

    \item \textbf{Exclusion Restriction}: The instrument $\mathbf{Z}$ should not have a direct effect on the outcome $\mathbf{Y}$, except through the exposure $\mathbf{X}$. This implies that the effect of $\mathbf{Z}$ on $\mathbf{Y}$ operates solely through $\mathbf{X}$, ensuring no other pathways influence the outcome. Mathematically, this is expressed as:
    \[
    \text{Cov}(\mathbf{Z}, \mathbf{Y} \mid \mathbf{X}) = 0.
    \]
\end{enumerate}
An instrument may have effects on multiple traits.
However, we require each trait to have at least one valid instrument for model identifiability. 



\color{black}

\section{Implemenation of RGM and NetworkMotif}

In this section, we consider illustrating some functionalities of the \textbf{MR.RGM} package, where we implement the primary functions \textbf{RGM} and \textbf{NetworkMotif} to some simulated examples.

\subsection{Installation of the Package from CRAN}

The following lines of \texttt{R} code shows the installation of the package from CRAN~\cite{CRAN}.

\begin{lstlisting}[language=R, numbers = none]
install.packages("MR.RGM")

# loading the package
library(MR.RGM)
\end{lstlisting}

\subsection{Simulating Data for Implementation}
We generate simulated data based on the assumed model and then proceed to demonstrate the implementation of \textbf{RGM} and \textbf{NetworkMotif}.

\begin{lstlisting}[language=R, numbers = none]

# Model: Y = AY + BX + E

# Set seed
set.seed(9154)

# Number of data points
n = 10000

# Number of response variables and number of instrument variables
p = 5
k = 6

# Initialize causal interaction matrix between response variables
A = matrix(sample(c(-0.1, 0.1), p^2, replace = TRUE), p, p)

# Diagonal entries of A matrix will always be 0
diag(A) = 0

# Make the network sparse
A[sample(which(A!=0), length(which(A!=0))/2)] = 0

# Create D matrix (Indicator matrix where each row corresponds to a response variable
# and each column corresponds to an instrument variable)
D = matrix(0, nrow = p, ncol = k)

# Manually assign values to D matrix
D[1, 1:2] = 1  # First response variable is influenced by the first 2 instruments
D[2, 3] = 1    # Second response variable is influenced by the 3rd instrument
D[3, 4] = 1    # Third response variable is influenced by the 4th instrument
D[4, 5] = 1    # Fourth response variable is influenced by the 5th instrument
D[5, 6] = 1    # Fifth response variable is influenced by the 6th instrument


# Initialize B matrix
B = matrix(0, p, k)  # Initialize B matrix with zeros

# Calculate B matrix based on D matrix
for (i in 1:p) {
   for (j in 1:k) {
     if (D[i, j] == 1) {
       B[i, j] = 1  # Set B[i, j] to 1 if D[i, j] is 1
     }
   }
 }



# Create variance-covariance matrix
Sigma = 1 * diag(p)

Mult_Mat = solve(diag(p) - A)

Variance = Mult_Mat %*% Sigma %*% t(Mult_Mat)

# Generate instrument data matrix
X = matrix(runif(n * k, 0, 5), nrow = n, ncol = k)

# Initialize response data matrix
Y = matrix(0, nrow = n, ncol = p)

# Generate response data matrix based on instrument data matrix
for (i in 1:n) {

 Y[i, ] = MASS::mvrnorm(n = 1, Mult_Mat %*% B %*% X[i, ], Variance)

}

# Print true causal interaction matrices between response variables and between response and instrument variables
A

#>      [,1] [,2] [,3] [,4] [,5]
#> [1,]  0.0 -0.1  0.0  0.0  0.1
#> [2,]  0.1  0.0 -0.1  0.1  0.1
#> [3,]  0.0 -0.1  0.0  0.0  0.1
#> [4,]  0.0 -0.1  0.0  0.0  0.0
#> [5,]  0.0  0.1  0.0  0.0  0.0

B

#>      [,1] [,2] [,3] [,4] [,5] [,6]
#> [1,]    1    1    0    0    0    0
#> [2,]    0    0    1    0    0    0
#> [3,]    0    0    0    1    0    0
#> [4,]    0    0    0    0    1    0
#> [5,]    0    0    0    0    0    1

\end{lstlisting}

\subsection{Using RGM Function in MR.RGM}

Initially, we demonstrate the application of the \textbf{RGM} function on the simulated dataset under varied prior assumptions, followed by a comparison of the graphical structures generated by \textbf{RGM} with the actual graph structures existing between response variables and between response and instrument variables.

\begin{lstlisting}[language = R, numbers = none]

# Apply RGM on individual level data with Threshold prior
Output1 = RGM(X = X, Y = Y, D = D, prior = "Threshold")

# Calculate summary level data
Syy = t(Y) %*% Y / n
Syx = t(Y) %*% X / n
Sxx = t(X) %*% X / n

# Apply RGM on summary level data for Spike and Slab Prior
Output2 = RGM(Syy = Syy, Syx = Syx, Sxx = Sxx, D = D, n = 10000, prior = "Spike and Slab")

# Calculate Beta and SigmaHat
# Centralize Data
Y = t(t(Y) - colMeans(Y))
X = t(t(X) - colMeans(X))

# Calculate Sxx
Sxx = t(X) %*% X / n

# Generate Beta matrix and SigmaHat
Beta = matrix(0, nrow = p, ncol = k)
SigmaHat = matrix(0, nrow = p, ncol = k)

for (i in 1:p) {

    for (j in 1:k) {

        fit = lm(Y[, i] ~ X[, j])

        Beta[i, j] =  fit$coefficients[2]

        SigmaHat[i, j] = sum(fit$residuals^2) / n

        }

 }


# Apply RGM on Sxx, Beta and SigmaHat for Spike and Slab Prior
Output3 = RGM(Sxx = Sxx, Beta = Beta, SigmaHat = SigmaHat,
           D = D, n = 10000, prior = "Spike and Slab")
           
\end{lstlisting}

The estimated causal interaction matrices between response variables are obtained as follows,

\begin{lstlisting}[language = R, numbers = none]

Output1$AEst
#>           [,1]        [,2]       [,3]      [,4]       [,5]
#> [1,] 0.0000000 -0.11032661  0.0000000 0.0000000 0.10676811
#> [2,] 0.0991208  0.00000000 -0.1104576 0.1002182 0.11012341
#> [3,] 0.0000000 -0.09370579  0.0000000 0.0000000 0.09747664
#> [4,] 0.0000000 -0.10185959  0.0000000 0.0000000 0.00000000
#> [5,] 0.0000000  0.10045256  0.0000000 0.0000000 0.00000000
Output2$AEst
#>              [,1]       [,2]          [,3]          [,4]        [,5]
#> [1,]  0.000000000 -0.1127249 -0.0006355133  0.0012237310 0.107520412
#> [2,]  0.099881480  0.0000000 -0.1079853541  0.0996589823 0.109275947
#> [3,] -0.001747246 -0.0929592  0.0000000000 -0.0006473297 0.099778267
#> [4,] -0.003863658 -0.1030056 -0.0024683725  0.0000000000 0.009191016
#> [5,]  0.001966818  0.1014136 -0.0055458038 -0.0050688662 0.000000000
Output3$AEst
#>             [,1]        [,2]         [,3]          [,4]       [,5]
#> [1,]  0.00000000 -0.08972628  0.039412442 -0.0006516754 0.09454632
#> [2,]  0.11040214  0.00000000 -0.112569739  0.0983975424 0.13676682
#> [3,]  0.01896308 -0.09780459  0.000000000  0.0100459671 0.11765744
#> [4,]  0.00000000 -0.11255315  0.001661851  0.0000000000 0.01358832
#> [5,] -0.00174675  0.12947084  0.011533919 -0.0025906139 0.00000000

\end{lstlisting}

The estimated causal network structures between the response variables are as follows,

\begin{lstlisting}[language = R, numbers = none]

Output1$zAEst
#>      [,1] [,2] [,3] [,4] [,5]
#> [1,]    0    1    0    0    1
#> [2,]    1    0    1    1    1
#> [3,]    0    1    0    0    1
#> [4,]    0    1    0    0    0
#> [5,]    0    1    0    0    0
Output2$zAEst
#>      [,1] [,2] [,3] [,4] [,5]
#> [1,]    0    1    0    0    1
#> [2,]    1    0    1    1    1
#> [3,]    0    1    0    0    1
#> [4,]    0    1    0    0    0
#> [5,]    0    1    0    0    0
Output3$zAEst
#>      [,1] [,2] [,3] [,4] [,5]
#> [1,]    0    1    0    0    1
#> [2,]    1    0    1    1    1
#> [3,]    0    1    0    0    1
#> [4,]    0    1    0    0    0
#> [5,]    0    1    0    0    0

\end{lstlisting}

We observe that the causal network structures inferred in the three outputs mentioned are identical. To gain a clearer understanding of the network, we compare the true network structure with the one estimated by \textbf{RGM}. Since the networks derived from all three outputs are consistent, we plot a single graph representing the estimated causal network.

\begin{lstlisting}[language = R, numbers = none]

# Define a function to create smaller arrowheads
smaller_arrowheads <- function(graph) {
  igraph::E(graph)$arrow.size = 0.60  # Adjust the arrow size value as needed
  return(graph)
}

# Create a layout for multiple plots
par(mfrow = c(1, 2))

# Plot the true causal network
plot(smaller_arrowheads(igraph::graph_from_adjacency_matrix((A != 0) * 1,
       mode = "directed")), layout = igraph::layout_in_circle,
          main = "True Causal Network")

# Plot the estimated causal network
plot(smaller_arrowheads(igraph::graph_from_adjacency_matrix(Output1$zAEst,
      mode = "directed")), layout = igraph::layout_in_circle,
         main = "Estimated Causal Network")
    
\end{lstlisting}

\begin{figure}[H]
    \centering
    \includegraphics[width = 15cm]{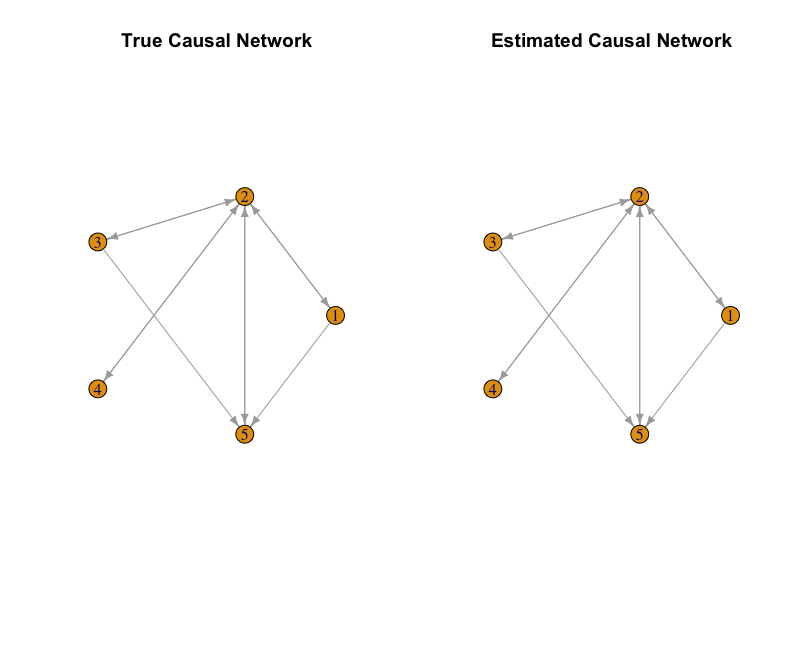}
    \caption{Comparison of the true causal network with the one generated by \textbf{RGM}.}
    \label{fig:1}
\end{figure}

We get the estimated causal interaction matrices between the response and the instrument variables from the outputs in the following way,

\begin{lstlisting}[language = R, numbers = none]

Output1$BEst
#>           [,1]     [,2]     [,3]      [,4]      [,5]      [,6]
#> [1,] 0.9935119 1.008009 0.000000 0.0000000 0.0000000 0.0000000
#> [2,] 0.0000000 0.000000 0.997496 0.0000000 0.0000000 0.0000000
#> [3,] 0.0000000 0.000000 0.000000 0.9998662 0.0000000 0.0000000
#> [4,] 0.0000000 0.000000 0.000000 0.0000000 0.9995511 0.0000000
#> [5,] 0.0000000 0.000000 0.000000 0.0000000 0.0000000 0.9982094
Output2$BEst
#>           [,1]   [,2]      [,3]      [,4]      [,5]     [,6]
#> [1,] 0.9937211 1.0075 0.0000000 0.0000000 0.0000000 0.000000
#> [2,] 0.0000000 0.0000 0.9964755 0.0000000 0.0000000 0.000000
#> [3,] 0.0000000 0.0000 0.0000000 0.9990566 0.0000000 0.000000
#> [4,] 0.0000000 0.0000 0.0000000 0.0000000 0.9987975 0.000000
#> [5,] 0.0000000 0.0000 0.0000000 0.0000000 0.0000000 1.002271
Output3$BEst
#>           [,1]     [,2]      [,3]      [,4]      [,5]      [,6]
#> [1,] 0.9902686 1.004035 0.0000000 0.0000000 0.0000000 0.0000000
#> [2,] 0.0000000 0.000000 0.9928588 0.0000000 0.0000000 0.0000000
#> [3,] 0.0000000 0.000000 0.0000000 0.9988565 0.0000000 0.0000000
#> [4,] 0.0000000 0.000000 0.0000000 0.0000000 0.9970532 0.0000000
#> [5,] 0.0000000 0.000000 0.0000000 0.0000000 0.0000000 0.9965687
    
\end{lstlisting}

The estimated graph structures between the response and the instrument variables are obtained as follows,

\begin{lstlisting}[language = R, numbers = none]

Output1$zBEst
#>      [,1] [,2] [,3] [,4] [,5] [,6]
#> [1,]    1    1    0    0    0    0
#> [2,]    0    0    1    0    0    0
#> [3,]    0    0    0    1    0    0
#> [4,]    0    0    0    0    1    0
#> [5,]    0    0    0    0    0    1
Output2$zBEst
#>      [,1] [,2] [,3] [,4] [,5] [,6]
#> [1,]    1    1    0    0    0    0
#> [2,]    0    0    1    0    0    0
#> [3,]    0    0    0    1    0    0
#> [4,]    0    0    0    0    1    0
#> [5,]    0    0    0    0    0    1
Output3$zBEst
#>      [,1] [,2] [,3] [,4] [,5] [,6]
#> [1,]    1    1    0    0    0    0
#> [2,]    0    0    1    0    0    0
#> [3,]    0    0    0    1    0    0
#> [4,]    0    0    0    0    1    0
#> [5,]    0    0    0    0    0    1

\end{lstlisting}

We can plot the log-likelihoods from the outputs in the following way,

\begin{lstlisting}[language = R, numbers = none]

plot(Output1$LLPst, type = 'l', xlab = "Iterations", ylab = "Log-likelihood")
    
\end{lstlisting}

\begin{figure}[H]
    \centering
    \includegraphics[width = 10cm]{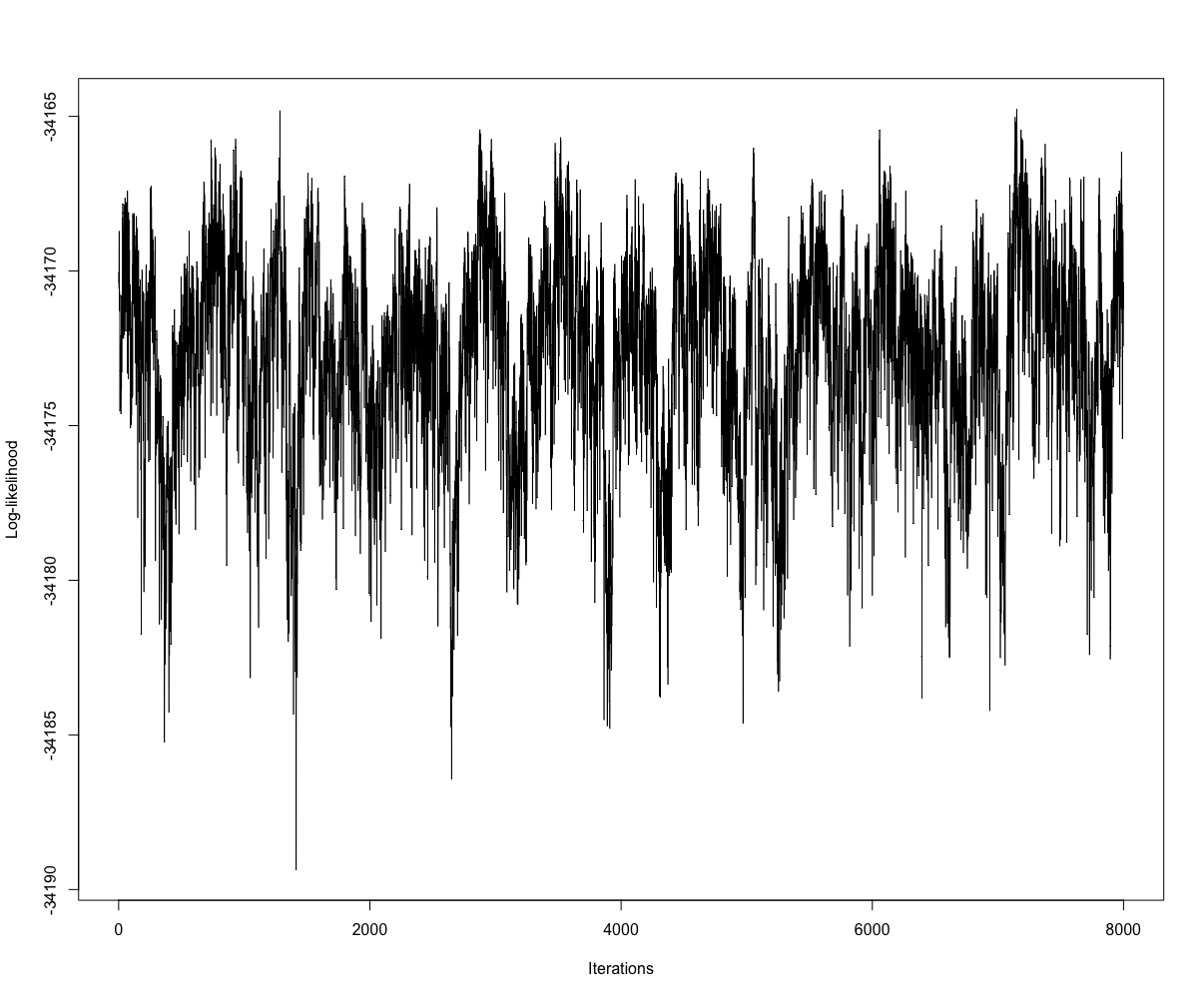}
    \caption{Posterior samples of log-likelihood vs Iterations.}
    \label{fig:2}
\end{figure}

\begin{lstlisting}[language = R, numbers = none]

plot(Output2$LLPst, type = 'l', xlab = "Iterations", ylab = "Log-likelihood")
    
\end{lstlisting}

\begin{figure}[H]
    \centering
    \includegraphics[width = 10cm]{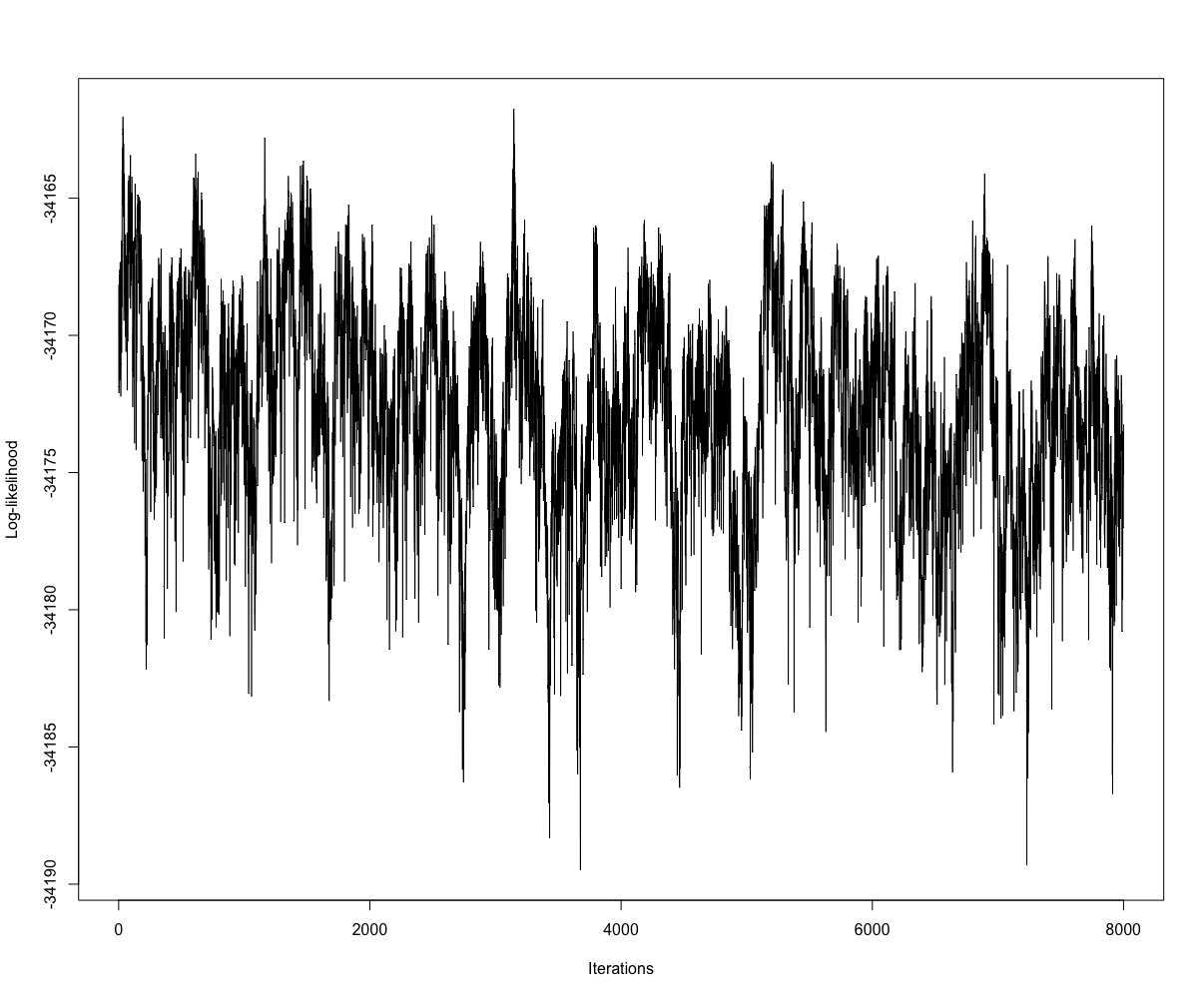}
    \caption{Posterior samples of log-likelihood vs Iterations.}
    \label{fig:3}
\end{figure}

\begin{lstlisting}[language = R, numbers = none]

plot(Output3$LLPst, type = 'l', xlab = "Iterations", ylab = "Log-likelihood")
    
\end{lstlisting}

\begin{figure}[H]
    \centering
    \includegraphics[width = 10cm]{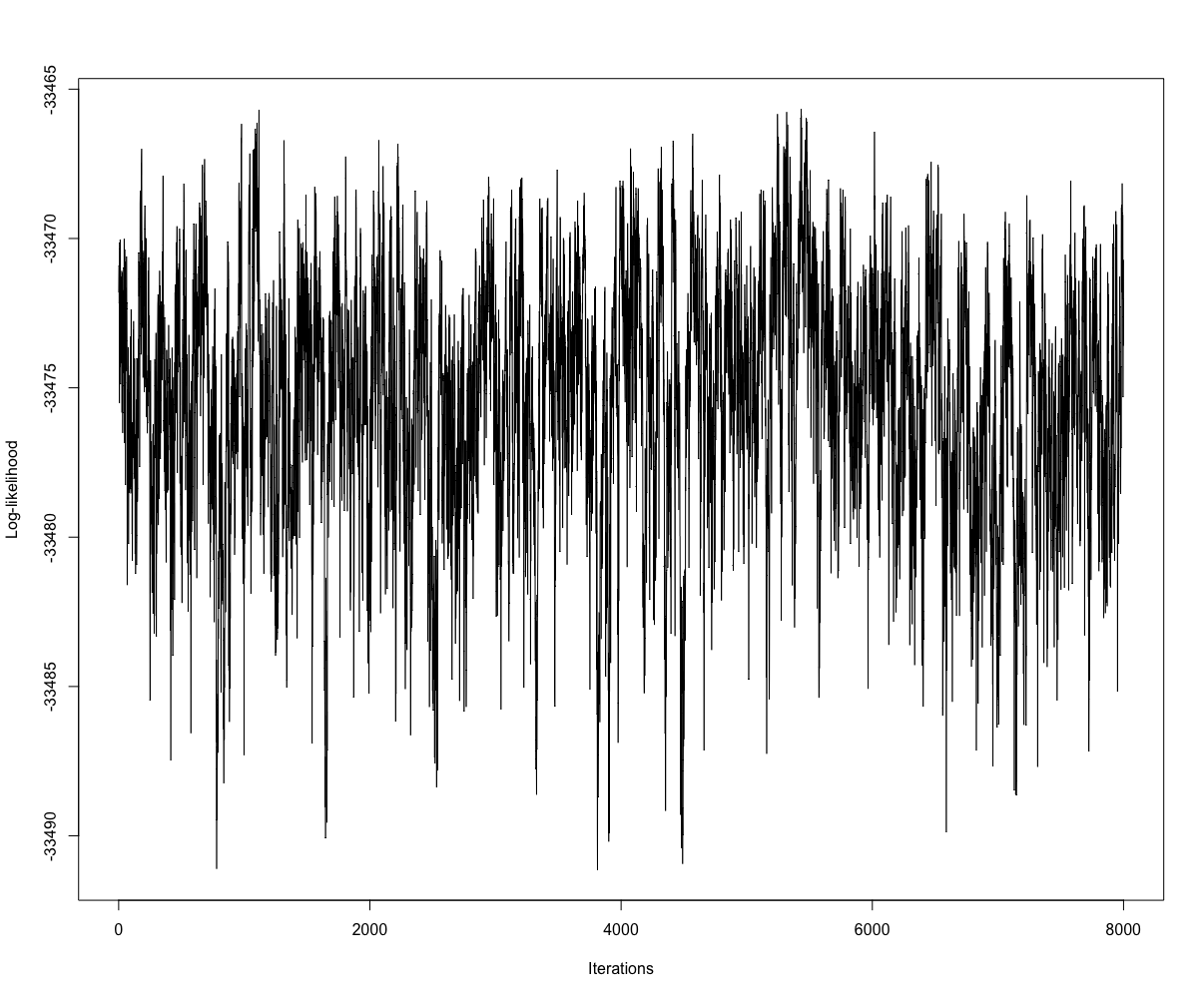}
    \caption{Posterior samples of log-likelihood vs Iterations.}
    \label{fig:3}
\end{figure}


\subsection{Expanded Simulation Setup}

In real-world scenarios, it is common to encounter a large number of instrumental variables (IVs), each explaining only a small proportion of the trait variance. To assess our method in such scenarios, we have expanded our earlier simulations with the following additional elements:

\subsubsection{Initial Setup with Numerous IVs}

We start with $100$ IVs for each response variable, each IV having a small effect on the trait. This setup mimics the situation where each IV contributes minimally to the overall trait variance, as often observed in real GWAS studies.

\subsubsection{Dimensionality Reduction}

To manage the high dimensionality of the data and focus on the most significant aspects of the variability, we perform PCA on each set of SNPs associated with a response variable. We select the top \texttt{20} principal components (PCs) from each PCA. This step reduces the complexity of the data while retaining the key features that explain the majority of the variance. By combining the top PCs from each response variable, we form a condensed matrix \texttt{compact\_X}. This matrix aggregates the instrumental variables into a more manageable form, facilitating a more efficient analysis.

\subsubsection{Revised Summary Level Data}

Using \texttt{compact\_X}, we calculate new summary level data (\texttt{Sxx\_compact} and \texttt{Syx\_compact}) for the RGM function application. 

\subsubsection{Updated R Code}

Here is the updated R code reflecting these changes:

\begin{lstlisting}[language = R, numbers = none]

# Load necessary libraries
library(MASS)
library(igraph)

# Set seed for reproducibility
set.seed(9154)

# Number of data points
n = 10000

# Number of response variables
p = 5

# Number of SNPs per response variable
num_snps_per_y = 100

# Total number of SNPs
k = num_snps_per_y * p

# Initialize causal interaction matrix between response variables
A = matrix(sample(c(-0.1, 0.1), p^2, replace = TRUE), p, p)
diag(A) = 0
A[sample(which(A != 0), length(which(A != 0)) / 2)] = 0

# Create D matrix (Indicator matrix where each row corresponds to a response variable
# and each column corresponds to an instrument variable)
D = matrix(0, nrow = p, ncol = k)

# Assign values to D matrix using a loop
for (run in 1:p) {
  D[run, ((run - 1) * 100 + 1):(run * 100)] = 1
}

# Initialize B matrix
B = matrix(0, p, k)  # Initialize B matrix with zeros

# Calculate B matrix based on D matrix
for (i in 1:p) {
  for (j in 1:k) {
    if (D[i, j] == 1) {
      B[i, j] = 1  # Set B[i, j] to 1 if D[i, j] is 1
    }
  }
}

# Calculate Variance-Covariance matrix
Sigma = diag(p)
Mult_Mat = solve(diag(p) - A)
Variance = Mult_Mat %*% Sigma %*% t(Mult_Mat)

# Generate instrument data matrix (X)
X = matrix(rnorm(n * k, 0, 1), nrow = n, ncol = k)

# Initialize response data matrix (Y)
Y = matrix(0, nrow = n, ncol = p)

# Generate response data matrix based on instrument data matrix
for (i in 1:n) {
  Y[i, ] = MASS::mvrnorm(n = 1, Mult_Mat %*% B %*% X[i, ], Variance)
}

# Calculate summary level data
Syy = t(Y) %*% Y / n
Syx = t(Y) %*% X / n
Sxx = t(X) %*% X / n

# Perform PCA for each response variable to get top 20 PCs
top_snps_list = list()
for (i in 1:p) {
  X_sub = X[, (num_snps_per_y * (i - 1) + 1):(num_snps_per_y * i)]
  pca = prcomp(X_sub, center = TRUE, scale. = TRUE)
  top_20_pcs = pca$x[, 1:20]
  top_snps_list[[i]] = top_20_pcs
}

# Combine the top PCs from all response variables
compact_X = do.call(cbind, top_snps_list)

# Calculate summary level data based on compact_X
Sxx_compact = t(compact_X) %*% compact_X / n
Syx_compact = t(Y) %*% compact_X / n

# Create D_New
D_New = matrix(0, nrow = p, ncol = 20 * p)

# Assign values to D matrix using a loop
for (run in 1:p) {
  D_New[run, ((run - 1) * 20 + 1):(run * 20)] = 1
}

# Apply RGM on summary level data for Spike and Slab Prior using the compact_X matrix
Output = RGM(Syy = Syy, Syx = Syx_compact, Sxx = Sxx_compact, D = D_New, n = n, prior = "Spike and Slab")

# Print estimated causal interaction matrices
Output$AEst
#>               [,1]        [,2]         [,3]         [,4]       [,5]
#> [1,]  0.0000000000 -0.10304239  0.012795897  0.006911765 0.09119095
#> [2,]  0.1102430697  0.00000000 -0.101718794  0.095093965 0.10762267
#> [3,] -0.0039579652 -0.10801784  0.000000000 -0.008388722 0.10255348
#> [4,] -0.0013919956 -0.09076408  0.011536478  0.000000000 0.01075220
#> [5,] -0.0005533104  0.10848253 -0.002160112 -0.011725145 0.00000000
Output$zAEst
#>      [,1] [,2] [,3] [,4] [,5]
#> [1,]    0    1    0    0    1
#> [2,]    1    0    1    1    1
#> [3,]    0    1    0    0    1
#> [4,]    0    1    0    0    0
#> [5,]    0    1    0    0    0

# Create a layout for multiple plots
par(mfrow = c(1, 2))

# Plot the true causal network
plot(smaller_arrowheads(igraph::graph_from_adjacency_matrix((A != 0) * 1, mode = "directed")), layout = igraph::layout_in_circle, main = "True Causal Network")

# Plot the estimated causal network
plot(smaller_arrowheads(igraph::graph_from_adjacency_matrix(Output$zAEst, mode = "directed")), layout = igraph::layout_in_circle, main = "Estimated Causal Network")

\end{lstlisting}

\begin{figure}[H]
    \centering
    \includegraphics[width = 15cm]{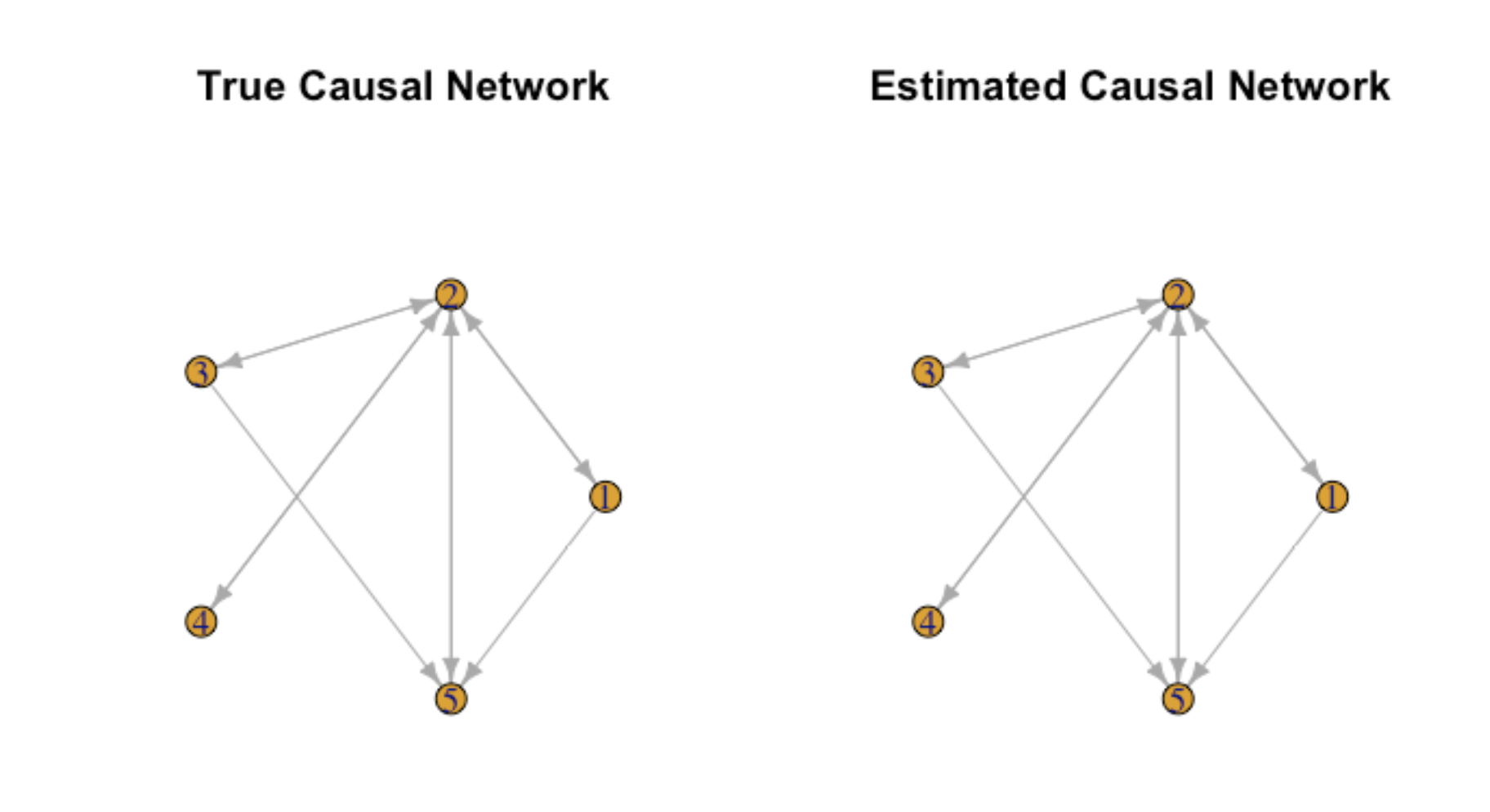  }
    \caption{Comparison of the true causal network with the one generated by \textbf{RGM}.}
    \label{fig:1}
\end{figure}

Our approach again yields very promising results, which demonstrates that our method could be useful in complex scenarios with many weak IVs.

The dimension reduction technique (PCA) proves to be effective. This approach can be broadly applied to similar problems where dimensionality reduction is necessary. By leveraging PCA or other dimension reduction methods, researchers can efficiently manage large sets of IVs and apply our algorithm to gain valuable insights into causal relationships.

\color{black}

\subsection{Using NetworkMotif Function in MR.RGM}

Next, we present the implementation of the \textbf{NetworkMotif} function. We begin by defining a random subgraph among the response variables. Subsequently, we collect $\textbf{GammaPst}$ arrays from various outputs and proceed to execute \textbf{NetworkMotif} based on these arrays.

\begin{lstlisting}[language = R, numbers = none]

# Start with a random subgraph
Gamma = matrix(0, nrow = p, ncol = p)
Gamma[5, 2] = Gamma[3, 5] = Gamma[2, 3] = 1

# Plot the subgraph to get an idea about the causal network
plot(smaller_arrowheads(igraph::graph_from_adjacency_matrix(Gamma,
       mode = "directed")), layout = igraph::layout_in_circle,
          main = "Subgraph")
    
\end{lstlisting}

\begin{figure}[H]
    \centering
    \includegraphics[width = 15cm]{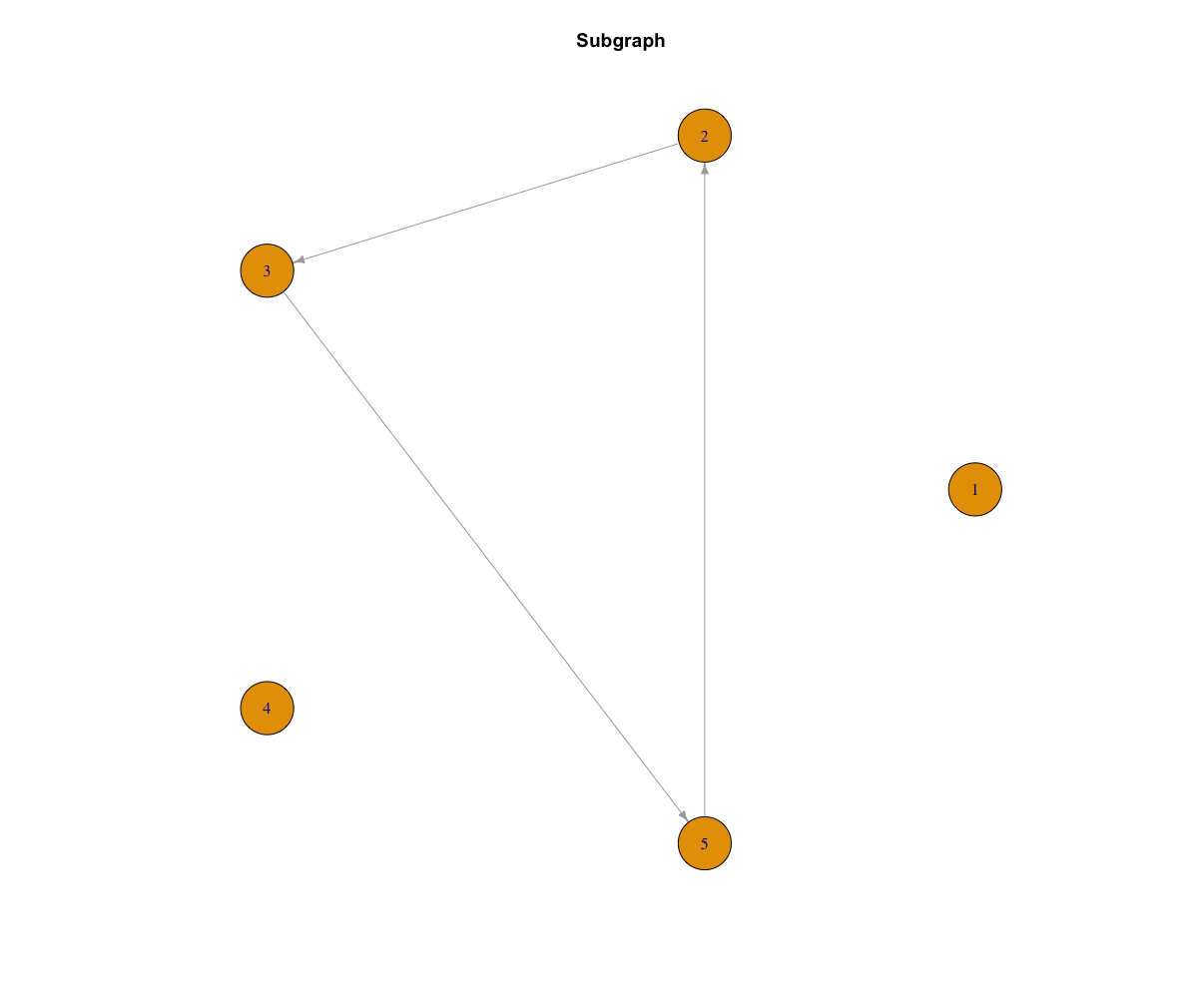}
    \caption{Subgraph between the response variables for which we want to caluclate posterior probability for.}
    \label{fig:4}
\end{figure}

\begin{lstlisting}[language = R, numbers = none]



# Store the GammaPst arrays from outputs
GammaPst1 = Output1$GammaPst
GammaPst2 = Output2$GammaPst
GammaPst3 = Output3$GammaPst

# Get the posterior probabilities of Gamma with these GammaPst matrices
NetworkMotif(Gamma = Gamma, GammaPst = GammaPst1)
#> [1] 1
NetworkMotif(Gamma = Gamma, GammaPst = GammaPst2)
#> [1] 0.3795
NetworkMotif(Gamma = Gamma, GammaPst = GammaPst3)
#> [1] 0.485125

\end{lstlisting}

\section{Computational complexity reduction for RGM}
\subsection{Calculation of det($\mathbf{I_p - A}$) and $(\mathbf{I_p - A})^{-1}$}

For MCMC sampling we have to calculate det($\mathbf{I_p - A}$) while sampling each entry of $\textbf{A}$. So, we have to calculate the determinant $p \times (p - 1)$ many times inside each iteration and if $p$ is large then this is heavily time consuming. In order to reduce complexity we have implemented woodbury matrix determinant lemma to calculate det($\mathbf{I_p - A}$). Woodbury matrix determinant lemma says:
\begin{align} \label{woodbury det}
    \text{det}(\mathbf{A+uv}^{T}) = (1 + \mathbf{v}^{T}\mathbf{A}^{-1}\mathbf{u})\text{det}(\mathbf{A})
\end{align}
Now suppose we want to sample $(i, j)$th entry of $\mathbf{A}$. Suppose, we get $\mathbf{A^{*}}$ by replacing $a_{ij}$ ($(i,j)$th entry of $\mathbf{A}$) with $a_{ij}^{*}$ and keeping all other entries same as $\mathbf{A}$. Now, we want to calculate det($\mathbf{I_p - A^{*}}$). Now instead of directly calculating this, we will try to calculate this from det($\mathbf{I_p - A}$) and in this way we will keep on updating this. Thus, we have:
\begin{align} \label{det(I - A)}
    \begin{split}
        \text{det}(\mathbf{I_p - A^{*}}) = \text{det}(\mathbf{I_p - A + A - A^{*}})
    \end{split}
\end{align}
Now the $(i,j)$th entry of ($\mathbf{A - A^{*}}$) is $(a_{ij} - a_{ij}^{*})$ and the rest are $0$. Thus, ($\mathbf{A - A^{*}}$) can be written as $\mathbf{uv}^{T}$, where $\mathbf{u}$ is a column vector whose $i$th entry is $(a_{ij} - a_{ij}^{*})$ and the rest are $0$ and $\mathbf{v}$ is a column vector whose $j$th entry is $1$ and the rest are $0$. Thus, from equation (\ref{det(I - A)}), we have:
\begin{align} \label{det(I - A_star)}
        \text{det}(\mathbf{I_p - A^{*}}) & =  \text{det}(\mathbf{I_p - A + A - A^{*}}) \nonumber \\
        & = \text{det}(\mathbf{I_p - A} + \mathbf{uv}^{T}) \nonumber \\
        & = (1 + \mathbf{v}^{T}\mathbf{(I_p - A)}^{-1}\mathbf{u}).\text{det}(\mathbf{I_p - A}) \nonumber \quad [\text{Using equation}\,(\ref{woodbury det})]\\
        & = (1 + \mathbf{(I_p - A)}^{-1}_{(j, i)} \times (a_{ij} - a_{ij}^{*})). \text{det}(\mathbf{I_p - A})
\end{align}
where, $\mathbf{(I_p - A)}^{-1}_{(j, i)}$ denotes the $(j,i)$th entry of $\mathbf{(I_p - A)}^{-1}$. Now, we want to calculate $\mathbf{(I_p - A^{*})}^{-1}$, but instead of calculating it directly we will calculate it from $\mathbf{(I_p - A)}^{-1}$. We will use woodbury matrix identity which states:
\begin{align} \label{woodbury inverse}
    (\mathbf{A + UCV})^{-1} = \mathbf{A}^{-1} - \mathbf{A}^{-1}\mathbf{U}(\mathbf{C}^{-1} + \mathbf{V}\mathbf{A}^{-1}\mathbf{U})^{-1}\mathbf{VA}^{-1}
\end{align}
Thus, we can calculate $\mathbf{(I_p - A^{*})}^{-1}$ in the following way:
\begin{align} \label{(I-A_star)_inv}
    \mathbf{(I_p - A^{*})}^{-1} & = \mathbf{(I_p - A + A - A^{*})}^{-1} \nonumber \\
    & = \mathbf{(I_p - A + \mathbf{uv}^{T})}^{-1} \nonumber\\
    & = \mathbf{(I_p - A)}^{-1} - \mathbf{(I_p - A)}^{-1}\mathbf{u}(1 + \mathbf{v}^{T}\mathbf{(I_p - A)}^{-1}\mathbf{u})^{-1}\mathbf{v}^{T}\mathbf{(I_p - A)}^{-1} \,[\text{Using equation} \, (\ref{woodbury inverse})] \nonumber\\
    & = \mathbf{(I_p - A)}^{-1} - \frac{a_{ij} - a_{ij}^{*}}{1+ (a_{ij} - a_{ij}^{*}). \mathbf{(I_p - A)}^{-1}_{(j, i)}}.\mathbf{(I_p - A)}^{-1}_{(., i)} \times \mathbf{(I_p - A)}^{-1}_{(j, .)}
\end{align}
where, $\mathbf{(I_p - A)}^{-1}_{(., i)}$ indicates $i$th coumn of $\mathbf{(I_p - A)}^{-1}$, $\mathbf{(I_p - A)}^{-1}_{(j, .)}$ indicates $j$th row of $\mathbf{(I_p - A)}^{-1}$ and $\mathbf{(I_p - A)}^{-1}_{(j, i)}$ indicates $(j,i)$th entry of $\mathbf{(I_p - A)}^{-1}$. So, we have implemented equation (\ref{det(I - A_star)}) to calculate det$\mathbf{(I_p - A^{*})}$ from 
det$\mathbf{(I_p - A)}$ and implemented equation (\ref{(I-A_star)_inv}) to calculate $\mathbf{(I_p - A^{*})}^{-1}$ from $\mathbf{(I_p - A)}^{-1}$ and we keep on updating them instead of directly calculating them everytime.
\subsection{Reduction in complexity for matrix multiplication}
For MCMC sampling we have to do some matrix multiplications and take trace while sampling each entry of $\textbf{A}$. The trace terms which can be found in Section (\ref{sec:Likelihood}), are as follows:
\begin{align*}
    \text{Trace}1 & = -n \times \text{trace}(\mathbf{Syy \times A^{T} \times \Sigma^{-1}})\\
    \text{Trace}2 & = -n \times \text{trace}(\mathbf{Syy \times \Sigma^{-1} \times A})\\
    \text{Trace}3 & = n \times \text{trace}(\mathbf{Syy \times A^{T} \times \Sigma^{-1} \times A})\\
    \text{Trace}4 & = 2n \times \text{trace}(\mathbf{Syx \times B^{T} \times \Sigma^{-1} \times A})\\
\end{align*}
where, $\mathbf{\Sigma} = \text{diag}(\sigma_1, \cdots, \sigma_p)$ and hence $\mathbf{\Sigma}^{-1} = \text{diag}(1/\sigma_1, \cdots, 1/\sigma_p)$. Now, suppose we want to sample $(i, j)$th entry of $\mathbf{A}$. We get $\mathbf{A^{*}}$ by replacing $a_{ij}$ ($(i,j)$th entry of $\mathbf{A}$) with $a_{ij}^{*}$ and keeping all other entries same as $\mathbf{A}$. Instead of calculating the trace values directly for $\mathbf{A}^{*}$, we will update it based on previous trace values in the following way:
\begin{align*}
    \text{Trace}1\_\text{New} & = \text{Trace}1 - n \times (a_{ij}^{*} - a_{ij}) \times \mathbf{\Sigma}^{-1}_{(i,i)} \times \mathbf{Syy}_{(i,j)}\\
    \text{Trace}2\_\text{New} & = \text{Trace}2 - n \times (a_{ij}^{*} - a_{ij}) \times \mathbf{\Sigma}^{-1}_{(i,i)} \times \mathbf{Syy}_{(j,i)}\\
    \text{Trace}3\_\text{New} & = \text{Trace}3 + n \times (a_{ij}^{*} - a_{ij}) \times \mathbf{\Sigma}^{-1}_{(i,i)} \times \left(\mathbf{A}_{(i,.)} \times \mathbf{Syy}_{(.,j)} + \mathbf{Syy}_{(j,.)} \times \mathbf{{A}^{*}}^{T}_{(i,.)}\right)\\
    \text{Trace}4\_\text{New} & = \text{Trace}4 + 2n \times (a_{ij}^{*} - a_{ij}) \times \mathbf{\Sigma}^{-1}_{(i,i)} \times \left(\mathbf{B}_{(i,.)} \times \mathbf{Syx}^{T}_{(j,.)}\right)\\
\end{align*}
where, for any matrix $\mathbf{A}$, $\mathbf{A}_{(i,i)}$ denotes $(i,i)$th element of $\mathbf{A}$, $\mathbf{A}_{(i,.)}$ denotes $i$th row of $\mathbf{A}$ and $\mathbf{A}_{(.,i)}$ denotes $i$th column of $\mathbf{A}$. Similarly, while sampling each entry of $\mathbf{B}$, we have to calculate the following trace values: 
\begin{align*}
    \text{Trace}5 & = -2n \times \text{trace}(\mathbf{Syx \times B^{T} \times \Sigma^{-1} \times (I_p - A)})\\
    \text{Trace}6 & = n \times \text{trace}(\mathbf{Sxx \times B^{T} \times \Sigma^{-1} \times B})\\
\end{align*}
Now, suppose we want to sample $(i, j)$th entry of $\mathbf{B}$. We get $\mathbf{B^{*}}$ by replacing $b_{ij}$ ($(i,j)$th entry of $\mathbf{B}$) with $b_{ij}^{*}$ and keeping all other entries same as $\mathbf{B}$. Instead of calculating the trace values directly for $\mathbf{B}^{*}$, we will update it based on previous trace values in the following way:
\begin{align*}
    \text{Trace}5\_\text{New} & = \text{Trace}5 - 2n \times (b_{ij}^{*} - b_{ij}) \times \mathbf{\Sigma}^{-1}_{(i,i)} \times (\mathbf{(I_p - A)}_{(i,.)} \times \mathbf{Syx}_{(.,j)})\\
    \text{Trace}6\_\text{New} & = \text{Trace}6 + n \times (b_{ij}^{*} - b_{ij}) \times \mathbf{\Sigma}^{-1}_{(i,i)} \times (\mathbf{B}_{(i,.)} \times \mathbf{Sxx}_{(.,j)} + \mathbf{Sxx}_{(j,.)} \times \mathbf{{B}^{*}}^{T}_{(i,.)})
\end{align*}


\section{Computational Scalability of MR.RGM}

\begin{figure}[H]
    \centering
    \begin{minipage}{0.48\textwidth}
        \centering
        \includegraphics[width=\textwidth]{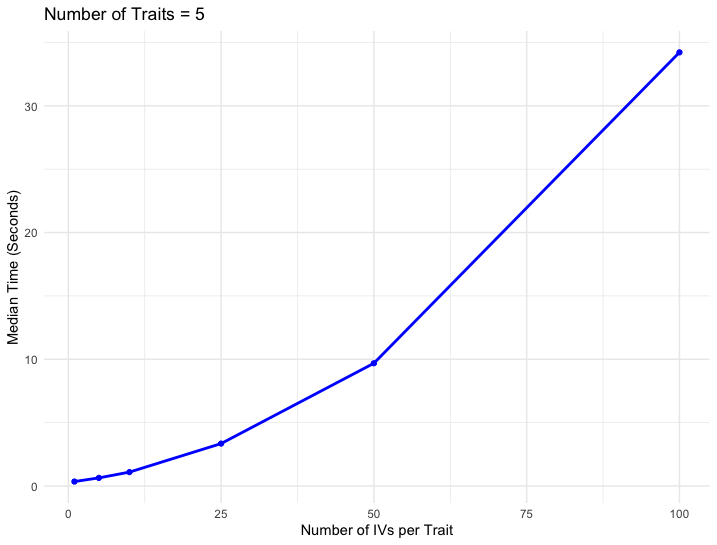}
        \caption{Median runtime for varying numbers of IVs per trait while fixing the number of traits at $5$.}
        \label{fig:setup1}
    \end{minipage}
    \hfill
    \begin{minipage}{0.48\textwidth}
        \centering
        \includegraphics[width=\textwidth]{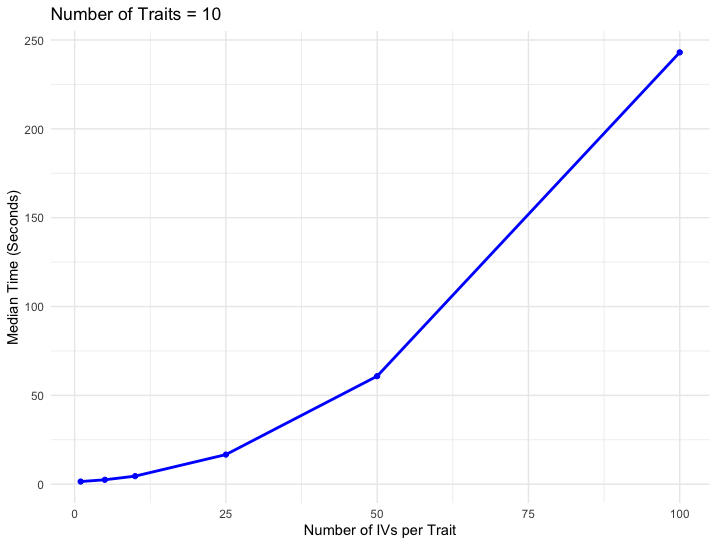}
        \caption{Median runtime for varying numbers of IVs per trait while fixing the number of traits at $10$.}
        \label{fig:setup2}
    \end{minipage}
\end{figure}

\begin{figure}[H]
    \centering
    \begin{minipage}{0.48\textwidth}
        \centering
        \includegraphics[width=\textwidth]{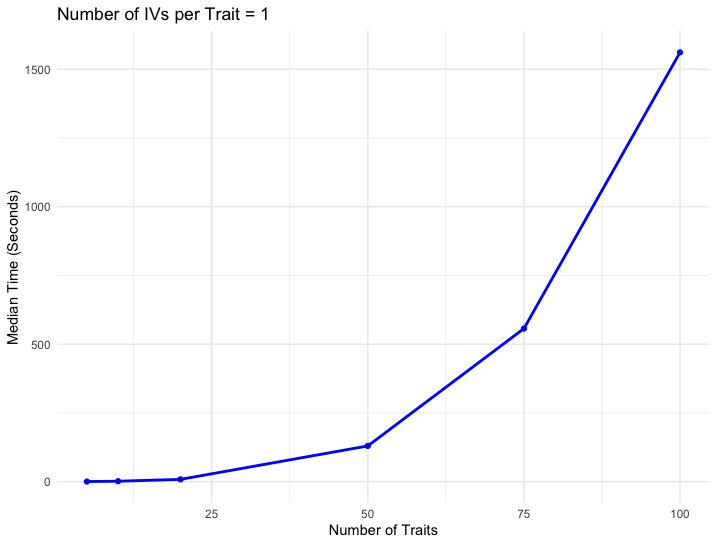}
        \caption{Median runtime for varying numbers of traits while fixing the number of IVs per trait at $1$.}
        \label{fig:setup3}
    \end{minipage}
    \hfill
    \begin{minipage}{0.48\textwidth}
        \centering
        \includegraphics[width=\textwidth]{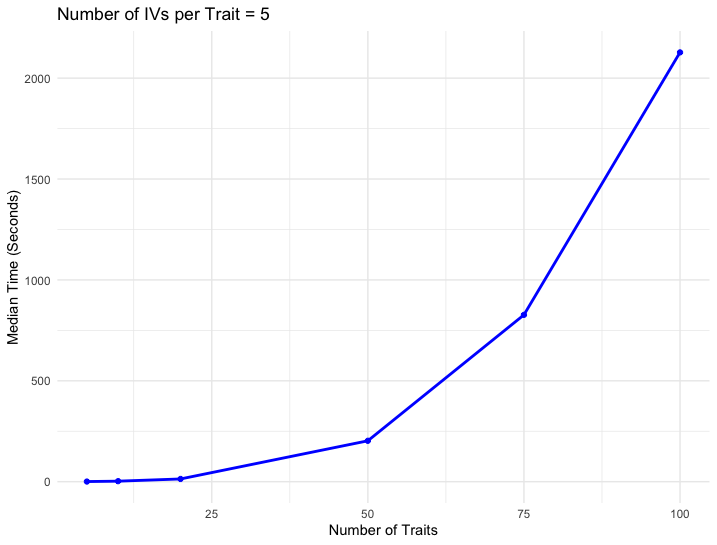}
        \caption{Median runtime for varying numbers of traits while fixing the number of IVs per trait at $5$.}
        \label{fig:setup4}
    \end{minipage}
\end{figure}

To evaluate the computational scalability of \textbf{MR.RGM}, we conducted a series of experiments to assess its performance with varying numbers of traits and instrumental variables (IVs). 
We considered four scenarios:

\begin{enumerate}
    \item \textbf{Fixing the number of traits at $5$ and varying the number of IVs per trait:} We tested with $1$, $5$, $10$, $25$, $50$, and $100$ IVs per trait. The results are illustrated in Figure~(\ref{fig:setup1}).
    \item \textbf{Fixing the number of traits at $10$ and varying the number of IVs per trait:} We used the same range of IVs as in the previous case. The corresponding results are shown in Figure~(\ref{fig:setup2}).
    \item \textbf{Fixing the number of IVs per trait at $5$ and varying the number of traits:} We tested with $5$, $10$, $20$, $50$, and $75$ traits. Results for this setup are presented in Figure~(\ref{fig:setup3}).
    \item \textbf{Fixing the number of IVs per trait at $10$ and varying the number of traits:} We used the same range of traits as in the previous case. These results are depicted in Figure~(\ref{fig:setup4}).
\end{enumerate}

In each setup, the method was run $100$ times, and for each run, \textbf{MR.RGM} performed $10,000$ MCMC iterations with ``Spike and Slab" prior. The median runtime was recorded.
All experiments were conducted on an Apple M$2$ Pro processor with 16 GB of memory.
\textbf{MR.RGM} demonstrated great computational efficiency compared to traditional Bayesian methods. \textbf{MR.RGM} scales relatively well with respect to the number of traits and IVs, as shown in the figures. This efficiency makes \textbf{MR.RGM} a practical choice for complex analyses involving numerous traits and instruments.

\noindent

\section{Sensitivity to the Assumption of Error Distributions}

To evaluate the sensitivity of \textbf{MR.RGM} to different error distributions, we conducted a comparative analysis with several error types:

\begin{itemize}
    \item \textbf{Normal Distribution:} Errors generated from normal distribution.
    \item \textbf{t-Distribution (df = $3$):} Errors generated from a t-distribution with $3$ degrees of freedom.
    \item \textbf{t-Distribution (df = $5$):} Errors generated from a t-distribution with $5$ degrees of freedom.
    \item \textbf{t-Distribution (df = $10$):} Errors generated from a t-distribution with $10$ degrees of freedom.
    \item \textbf{Laplace Distribution:} Errors generated from a Laplace distribution.
\end{itemize}

For each error distribution, \textbf{MR.RGM} was applied to the datasets, and the following metrics were computed:
\begin{itemize}
    \item \textbf{True Positive Rate (TPR):} The proportion of true edges correctly identified by \textbf{MR.RGM} in the estimated network structure.
    \item \textbf{False Positive Rate (FPR):} The proportion of false edges incorrectly identified in the estimated network structure.
    \item \textbf{Mean Absolute Deviation (MAD):} The average deviation between the estimated causal effects and the true causal effects.
\end{itemize}
These metrics were calculated based on the estimated network structure and causal effects, with the process repeated $1,000$ times for each distribution.

\subsection{Results}

Table~(\ref{tab:error_distributions}) summarizes the average TPR, FPR, and MAD for each error distribution. The results indicate that \textbf{MR.RGM} is not sensitive to the assumption of normal errors, as performance metrics are consistent across different error distributions. This suggests that \textbf{MR.RGM} performs robustly even when the error distribution deviates from normality.

\begin{table}[H]
    \centering
    \caption{Performance of \textbf{MR.RGM} under different error distributions.}
    \label{tab:error_distributions}
    \begin{tabular}{|c|c|c|c|}
        \hline
        \textbf{Error Distribution} & \textbf{TPR} & \textbf{FPR} & \textbf{MAD} \\
        \hline
        Normal Distribution & $0.86$ & $0.0133$ & $0.0169$ \\
        t-Distribution (df = $3$) & $0.86$ & $0.0200$ & $0.0153$ \\
        t-Distribution (df = $5$) & $0.92$ & $0.0400$ & $0.0154$ \\
        t-Distribution (df = $10$) & $0.86$ & $0.0467$ & $0.0179$ \\
        Laplace Distribution & $0.87$ & $0.0333$ & $0.0164$ \\
        \hline
    \end{tabular}
\end{table}

\color{black}

\section{Likelihood of the model based on summary level data}
\label{sec:Likelihood}

Now, we will illustrate the likelihood calculation when summary-level data is provided. The likelihood of the model, based on individual-level data, is expressed as follows:

\begin{align*}
    \begin{split}
        & p\left(\{\mathbf{Y_i}\}_{i=1}^n|\{\mathbf{X_i}\}_{i=1}^n, \mathbf{A, B, \Sigma}\right)\\
        = & \prod_{i=1}^nN(\mathbf{Y_i|(I_p-A)^{-1}BX_i,(I_p-A)^{-1} \Sigma (I_p - A)^{-T}})\\
        = & \prod_{i=1}^nN(\mathbf{(I_p - A)Y_i - BX_i | 0, \Sigma).|\text{det}(I_p - A)|}\\
        = & \prod_{i = 1}^n (2\pi)^{-p/2}\mathbf{\text{det}(\Sigma)^{-1/2}|\text{det}(I_p - A)|} \exp(-\frac{1}{2}\mathbf{[(I_p - A)Y_i - BX_i]^{T}\Sigma^{-1}[(I_p - A)Y_i - BX_i])}\\
        = & (2\pi)^{-\frac{np}{2}}.\mathbf{\text{det}(\Sigma)^{-\frac{n}{2}}.|\text{det}(I_p - A)|^n}.\exp(-\frac{1}{2}\sum_{i = 1}^n \mathbf{[(I_p - A)Y_i - BX_i]^{T}\Sigma^{-1}[(I_p - A)Y_i - BX_i])}\\
        = & (2\pi)^{-\frac{np}{2}}.\mathbf{\text{det}(\Sigma)^{-\frac{n}{2}}.|\text{det}(I_p - A)|^n}.\exp(-\frac{1}{2}\sum_{i = 1}^n \mathbf{[Y_i^{T}(I_p - A)^{T}\Sigma^{-1}(I_p - A)Y_i} \\
        & \quad \quad \quad \quad - 2\mathbf{X_i^{T}B^{T}\Sigma^{-1}(I_p - A)Y_i + X_i^{T}B^{T}\Sigma^{-1}BX_i])}\\
        = & (2\pi)^{-\frac{np}{2}}.\mathbf{\text{det}(\Sigma)^{-\frac{n}{2}}.|\text{det}(I_p - A)|^n}.\exp(-\frac{1}{2}\sum_{i = 1}^n \mathbf{[\text{tr}(Y_i^{T}(I_p - A)^{T}\Sigma^{-1}(I_p - A)Y_i)} \\
        & \quad \quad \quad \quad - 2\mathbf{\text{tr}(X_i^{T}B^{T}\Sigma^{-1}(I_p - A)Y_i) + \text{tr}(X_i^{T}B^{T}\Sigma^{-1}BX_i)]})\\
        = & (2\pi)^{-\frac{np}{2}}.\mathbf{\text{det}(\Sigma)^{-\frac{n}{2}}.|\text{det}(I_p - A)|^n}.\exp(-\frac{1}{2}\sum_{i = 1}^n \mathbf{[\text{tr}(Y_iY_i^{T}(I_p - A)^{T}\Sigma^{-1}(I_p - A))} \\
        & \quad \quad \quad \quad - 2\mathbf{\text{tr}(Y_iX_i^{T}B^{T}\Sigma^{-1}(I_p - A)) + \text{tr}(X_iX_i^{T}B^{T}\Sigma^{-1}B)]})\\
    \end{split}
\end{align*}


\begin{align*}
    \begin{split}
        = & (2\pi)^{-\frac{np}{2}}.\mathbf{\text{det}(\Sigma)^{-\frac{n}{2}}.|\text{det}(I_p - A)|^n}.\exp(-\frac{1}{2} [\text{tr}(\sum_{i = 1}^n\mathbf{Y_iY_i^{T}(I_p - A)^{T}\Sigma^{-1}(I_p - A)}) \\
        & \quad \quad \quad \quad - 2\text{tr}(\sum_{i = 1}^n\mathbf{Y_iX_i^{T}B^{T}\Sigma^{-1}(I_p - A)}) + \text{tr}(\sum_{i = 1}^n\mathbf{X_iX_i^{T}B^{T}\Sigma^{-1}B})])\\
        = & (2\pi)^{-\frac{np}{2}}.\mathbf{\text{det}(\Sigma)^{-\frac{n}{2}}.|\text{det}(I_p - A)|^n}.\exp(-\frac{1}{2} [\text{tr}(n.\mathbf{Syy (I_p - A)^{T}\Sigma^{-1}(I_p - A)}) \\
        & \quad \quad \quad \quad - 2\text{tr}(n.\mathbf{SyxB^{T}\Sigma^{-1}(I_p - A)}) + \text{tr}(n.\mathbf{SxxB^{T}\Sigma^{-1}B)}])\\
         = & (2\pi)^{-\frac{np}{2}}.\mathbf{\text{det}(\Sigma)^{-\frac{n}{2}}.|\text{det}(I_p - A)|^n}.\exp(-\frac{1}{2} [n.\text{tr}\mathbf{(Syy(I_p - A)^{T}\Sigma^{-1}(I_p - A))} \\
        & \quad \quad \quad \quad - 2n.\text{tr}\mathbf{(SyxB^{T}\Sigma^{-1}(I_p - A))} + n.\text{tr}\mathbf{(SxxB^{T}\Sigma^{-1}B)}])\\
    \end{split}
\end{align*}

\section{Calculation of $\textbf{A}-$matrix from $\textbf{Beta}-$matrix} 
\label{sec:Calculation of A matrix from Beta matrix}
Let $\mathbf{Y_i} = (Y_{i1}, \cdots, Y_{ip})^{T}$ denote the expressions for response variables $1, \cdots, p$, let $\mathbf{X_i} = (X_{i1}, \cdots, X_{ik})^{T}$ be the set of measurements for instrument variables $1, \cdots, k$ and for $i = 1, \ldots, n$. For this moment lets assume each response is affected by one and only one instrument i.e. $k = p$ and hence $\mathbf{B}$ is a diagonal matrix. We have the following model:
\begin{align} \label{Eq 7.1}
    & \mathbf{Y_i} = \mathbf{AY_i} + \mathbf{BX_i} + \mathbf{E_i}, \quad \mathbf{E_i} \sim N(0, \mathbf{\Sigma}) \nonumber\\
    \implies & (\mathbf{I_p} - \mathbf{A}) \mathbf{Y_i} = \mathbf{BX_i} + \mathbf{E_i} \nonumber\\
    \implies & \mathbf{Y_i} = (\mathbf{I_p} - \mathbf{A})^{-1}\mathbf{BX_i} + (\mathbf{I_p} - \mathbf{A})^{-1} \mathbf{E_i}   
\end{align}
where,
\begin{align*}
    \mathbf{A} = \begin{bmatrix} 
    0 & a_{12} & \dots  & a_{1p}\\
    a_{21} & 0 & \dots  & a_{2p}\\
    \vdots & \vdots & \ddots & \vdots\\
    a_{p1} & a_{p2} & \dots  & 0 
    \end{bmatrix} \quad
    \mathbf{B} = \begin{bmatrix} 
    b_{1} & 0 & \dots  & 0\\
    0 & b_{2} & \dots  & 0\\
    \vdots & \vdots & \ddots & \vdots\\
    0 & 0 & \dots  & b_{p} 
    \end{bmatrix}
\end{align*}
We define 
\begin{align*}
    \textbf{Beta} = \begin{bmatrix} 
    \hat\beta_{11} & \hat\beta_{12} & \dots  & \hat\beta_{1p}\\
    \hat\beta_{21} & \hat\beta_{22} & \dots  & \hat\beta_{2p}\\
    \vdots & \vdots & \ddots & \vdots\\
    \hat\beta_{p1} & \hat\beta_{p2} & \dots  & \hat\beta_{pp} 
    \end{bmatrix}
\end{align*}
where, $\hat\beta_{ij}$ denotes the regression coefficient of $Y_i$ on $X_j$. So, we can assume from equation (\ref{Eq 7.1}),
\begin{align} \label{beta}
    \textbf{Beta} = \mathbf{(I_p - A)}^{-1}\mathbf{B}
\end{align}
\textbf{Claim:} We claim that $a_{ij}$ can be calculated from the $\textbf{Beta}$ matrix in the following way:
\begin{align} \label{a_ij}
\begin{split}
    a_{ij} = & (-1)^{(i-j+1)} \times \frac{\operatorname{det}\left(\begin{bmatrix} 
    \hat\beta_{11} & \dots & \hat\beta_{1(i-1)} & \hat\beta_{1(i+1)} & \dots  & \hat\beta_{1p}\\
    \vdots & \ddots & \vdots & \vdots & \ddots & \vdots\\
    \hat\beta_{(j-1)1} & \dots & \hat\beta_{(j-1)(i-1)} & \hat\beta_{(j-1)(i+1)} & \dots  & \hat\beta_{(j-1)p}\\
    \hat\beta_{(j+1)1} & \dots & \hat\beta_{(j+1)(i-1)} & \hat\beta_{(j+1)(i+1)} & \dots  & \hat\beta_{(j+1)p}\\
    \vdots & \ddots & \vdots & \vdots & \ddots & \vdots\\
    \hat\beta_{p1} & \dots & \hat\beta_{p(i-1)} & \hat\beta_{p(i+1)} & \dots  & \hat\beta_{pp}\\
    \end{bmatrix}\right)}{\operatorname{det}\left(\begin{bmatrix} 
    \hat\beta_{11} & \dots & \hat\beta_{1(i-1)} & \hat\beta_{1(i+1)} & \dots  & \hat\beta_{1p}\\
    \vdots & \ddots & \vdots & \vdots & \ddots & \vdots\\
    \hat\beta_{(i-1)1} & \dots & \hat\beta_{(i-1)(i-1)} & \hat\beta_{(i-1)(i+1)} & \dots  & \hat\beta_{(i-1)p}\\
    \hat\beta_{(i+1)1} & \dots & \hat\beta_{(i+1)(i-1)} & \hat\beta_{(i+1)(i+1)} & \dots  & \hat\beta_{(i+1)p}\\
    \vdots & \ddots & \vdots & \vdots & \ddots & \vdots\\
    \hat\beta_{p1} & \dots & \hat\beta_{p(i-1)} & \hat\beta_{p(i+1)} & \dots  & \hat\beta_{pp} \\
    \end{bmatrix}\right)}
\end{split}
\end{align}
i.e. starting with $\textbf{Beta}$ matrix we first remove the $i$-th column, then from that matrix we first remove the $j$-th row and calculate the determinant and divide it by the determinant of the matrix calculated by removing the $i$-th row. The sign of $a_{ij}$ is determined by $(-1)^{i-j+1}$. We will now prove this.
Let,
\begin{align*}
    \mathbf{(I_p - A)}^{-1} = \begin{bmatrix} 
    M_{11} & M_{12} & \dots  & M_{1p}\\
    M_{21} & M_{22} & \dots  & M_{2p}\\
    \vdots & \vdots & \ddots & \vdots\\
    M_{p1} & M_{p2} & \dots  & M_{pp} 
    \end{bmatrix}
\end{align*}
From equation (\ref{beta}) we have,
\begin{align*}
   & \textbf{Beta} = \mathbf{(I_p - A)}^{-1}\mathbf{B}\\
   \implies & \begin{bmatrix} 
    \hat\beta_{11} & \hat\beta_{12} & \dots  & \hat\beta_{1p}\\
    \hat\beta_{21} & \hat\beta_{22} & \dots  & \hat\beta_{2p}\\
    \vdots & \vdots & \ddots & \vdots\\
    \hat\beta_{p1} & \hat\beta_{p2} & \dots  & \hat\beta_{pp} 
    \end{bmatrix} = \begin{bmatrix} 
    M_{11} & M_{12} & \dots  & M_{1p}\\
    M_{21} & M_{22} & \dots  & M_{2p}\\
    \vdots & \vdots & \ddots & \vdots\\
    M_{p1} & M_{p2} & \dots  & M_{pp} 
    \end{bmatrix} \times  \begin{bmatrix} 
    b_{1} & 0 & \dots  & 0\\
    0 & b_{2} & \dots  & 0\\
    \vdots & \vdots & \ddots & \vdots\\
    0 & 0 & \dots  & b_{p} 
    \end{bmatrix} \\
    \implies & \hat\beta_{ij} = M_{ij}b_{j} \quad \forall i,j \in \{1,\cdots,p\} 
\end{align*}
Thus replacing $\hat\beta_{ij}$ by $M_{ij}\times b_{j}$ in equation \eqref{a_ij} we get:
\begin{align*} 
\begin{split}
    a_{ij} = & (-1)^{(i-j+1)} \times \frac{\operatorname{det}\left(\begin{bmatrix} 
    M_{11}b_1 & \dots & M_{1(i-1)}b_{i-1} & M_{1(i+1)}b_{i+1} & \dots  & M_{1p}b_p\\
    \vdots & \ddots & \vdots & \vdots & \ddots & \vdots\\
    M_{(j-1)1}b_1 & \dots & M_{(j-1)(i-1)}b_{i-1} & M_{(j-1)(i+1)}b_{i + 1} & \dots  & M_{(j-1)p}b_p\\
    M_{(j+1)1}b_1 & \dots & M_{(j+1)(i-1)}b_{i-1} & M_{(j+1)(i+1)}b_{i+1} & \dots  & M_{(j+1)p}b_p\\
    \vdots & \ddots & \vdots & \vdots & \ddots & \vdots\\
    M_{p1}b_1 & \dots & M_{p(i-1)}b_{i-1} & M_{p(i+1)}b_{i+1} & \dots  & M_{pp}b_p \\
    \end{bmatrix}\right)}{\operatorname{det}\left(\begin{bmatrix} 
    M_{11}b_1 & \dots & M_{1(i-1)}b_{i-1} & M_{1(i+1)}b_{i + 1} & \dots  & M_{1p}b_p\\
    \vdots & \ddots & \vdots & \vdots & \ddots & \vdots\\
    M_{(i-1)1}b_1 & \dots & M_{(i-1)(i-1)}b_{i-1} & M_{(i-1)(i+1)}b_{i + 1} & \dots  & M_{(i-1)p}b_p\\
    M_{(i+1)1}b_1 & \dots & M_{(i+1)(i-1)}b_{i-1} & M_{(i+1)(i+1)}b_{i + 1} & \dots  & M_{(i+1)p}b_p\\
    \vdots & \ddots & \vdots & \vdots & \ddots & \vdots\\
    M_{p1}b_1 & \dots & M_{p(i-1)}b_{i-1} & M_{p(i+1)}b_{i + 1} & \dots  & M_{pp}b_p \\
    \end{bmatrix}\right)}
\end{split}
\end{align*}
\begin{align} \label{a_ij_new}
\begin{split}
    = & (-1)^{(i-j+1)} \times \frac{\operatorname{det}\left(\begin{bmatrix} 
    M_{11} & \dots & M_{1(i-1)} & M_{1(i+1)} & \dots  & M_{1p}\\
    \vdots & \ddots & \vdots & \vdots & \ddots & \vdots\\
    M_{(j-1)1} & \dots & M_{(j-1)(i-1)} & M_{(j-1)(i+1)} & \dots  & M_{(j-1)p}\\
    M_{(j+1)1} & \dots & M_{(j+1)(i-1)} & M_{(j+1)(i+1)} & \dots  & M_{(j+1)p}\\
    \vdots & \ddots & \vdots & \vdots & \ddots & \vdots\\
    M_{p1} & \dots & M_{p(i-1)} & M_{p(i+1)} & \dots  & M_{pp} \\
    \end{bmatrix}\right)}{\operatorname{det}\left(\begin{bmatrix} 
    M_{11} & \dots & M_{1(i-1)} & M_{1(i+1)} & \dots  & M_{1p}\\
    \vdots & \ddots & \vdots & \vdots & \ddots & \vdots\\
    M_{(i-1)1} & \dots & M_{(i-1)(i-1)} & M_{(i-1)(i+1)} & \dots  & M_{(i-1)p}\\
    M_{(i+1)1} & \dots & M_{(i+1)(i-1)} & M_{(i+1)(i+1)} & \dots  & M_{(i+1)p}\\
    \vdots & \ddots & \vdots & \vdots & \ddots & \vdots\\
    M_{p1} & \dots & M_{p(i-1)} & M_{p(i+1)} & \dots  & M_{pp} \\
    \end{bmatrix}\right)}
\end{split}
\end{align}
Again, we have, 
\begin{align*}
    & \mathbf{(I_p - A)} \times \mathbf{(I_p - A)}^{-1} = \mathbf{I_p}\\
    \implies & \begin{bmatrix} 
    1 & -a_{12} & \dots  & -a_{1p}\\
    -a_{21} & 1 & \dots  & -a_{2p}\\
    \vdots & \vdots & \ddots & \vdots\\
    -a_{p1} & -a_{p2} & \dots  & 1 
    \end{bmatrix} \times
    \begin{bmatrix} 
    M_{11} & M_{12} & \dots  & M_{1p}\\
    M_{21} & M_{22} & \dots  & M_{2p}\\
    \vdots & \vdots & \ddots & \vdots\\
    M_{p1} & M_{p2} & \dots  & M_{pp} 
    \end{bmatrix} =
    \begin{bmatrix} 
    1 & 0 & \dots  & 0\\
    0 & 1 & \dots  & 0\\
    \vdots & \vdots & \ddots & \vdots\\
    0 & 0 & \dots  & 1 
    \end{bmatrix}
\end{align*}
So, for each $i\in\{1,\cdots,p\}$ there will be $p$ many linear equations involving $a_{i1},\cdots,a_{i(i-1)},a_{i(i+1)},\cdots,a_{ip}$. We have the linear equations as follows:
\begin{align*}
     -a_{i1}M_{11}-a_{i2}M_{21}-\cdots+& M_{i1}-\cdots-a_{ip}M_{p1} = 0\\
     -a_{i1}M_{12}-a_{i2}M_{22}-\cdots+&M_{i2}-\cdots-a_{ip}M_{p2} = 0\\
    & \vdots \\
    -a_{i1}M_{1i}-a_{i2}M_{2i}-\cdots+&M_{ii}-\cdots-a_{ip}M_{pi} = 1\\
    & \vdots \\
    -a_{i1}M_{1p}-a_{i2}M_{2p}-\cdots+&M_{ip}-\cdots-a_{ip}M_{pp} = 0\\
\end{align*}
So, we have $(p-1)$ many variables $a_{i1},\cdots,a_{i(i-1)},a_{i(i+1)},\cdots,a_{ip}$ and $p$ many equations and we know that solutions for $a_{i1},\cdots,a_{i(i-1)},a_{i(i+1)},\cdots,a_{ip}$ exist which satisfy these $p$ many equations. So, we can take any $(p-1)$ many equations and solve them. Let us take the equations whose RHS is 0. So, we have:  
\begin{align*}
     -a_{i1}M_{11}-a_{i2}M_{21}-\cdots-&a_{ij} M_{j1}-\cdots-a_{ip}M_{p1} = -M_{i1}\\
     -a_{i1}M_{12}-a_{i2}M_{22}-\cdots-&a_{ij} M_{j2}-\cdots-a_{ip}M_{p2} = -M_{i2}\\
    & \vdots \\
    -a_{i1}M_{1(i-1)}-a_{i2}M_{2(i-1)}-\cdots-&a_{ij} M_{j(i-1)}-\cdots-a_{ip}M_{p(i-1)} = -M_{i(i-1)}\\
    -a_{i1}M_{1(i+1)}-a_{i2}M_{2(i+1)}-\cdots-&a_{ij} M_{j(i+1)}-\cdots-a_{ip}M_{p(i+1)} = -M_{i(i+1)}\\
    & \vdots\\
    -a_{i1}M_{1p}-a_{i2}M_{2p}-\cdots-&a_{ij} M_{jp}-\cdots-a_{ip}M_{pp} = -M_{ip}\\
\end{align*}
Thus, applying Cramer's rule to solve $(p-1)$ many linear equations in $(p-1)$ many unknowns we have the solutions as follows:

\begin{align*}
\begin{split}
    a_{ij} = & \frac{\operatorname{det}\left(\begin{bmatrix} 
    - M_{11} & \dots & M_{(j-1)1} & -M_{i1} & - M_{(j+1)1} & \dots  & -M_{p1}\\
    \vdots & \ddots & \vdots & \vdots & \vdots & \ddots & \vdots\\
    - M_{1(i-1)} & \dots & M_{(j-1)(i-1)} & -M_{i(i-1)} & - M_{(j+1)(i-1)} & \dots  & -M_{p(i-1)}\\
    - M_{1(i+1)} & \dots & M_{(j-1)(i+1)} & -M_{i(i+1)} & - M_{(j+1)(i+1)} & \dots  & -M_{p(i+1)}\\
      \vdots & \ddots & \vdots & \vdots & \vdots & \ddots & \vdots\\
    - M_{1p} & \dots & M_{(j-1)p} & -M_{ip} & - M_{(j+1)p} & \dots  & -M_{pp}\\
    \end{bmatrix}\right)}{\operatorname{det}\left(\begin{bmatrix} 
    - M_{11} & \dots & M_{(j-1)1} & -M_{j1} & - M_{(j+1)1} & \dots  & -M_{p1}\\
    \vdots & \ddots & \vdots & \vdots & \vdots & \ddots & \vdots\\
    - M_{1(i-1)} & \dots & M_{(j-1)(i-1)} & -M_{j(i-1)} & - M_{(j+1)(i-1)} & \dots  & -M_{p(i-1)}\\
    - M_{1(i+1)} & \dots & M_{(j-1)(i+1)} & -M_{j(i+1)} & - M_{(j+1)(i+1)} & \dots  & -M_{p(i+1)}\\
      \vdots & \ddots & \vdots & \vdots & \vdots & \ddots & \vdots\\
    - M_{1p} & \dots & M_{(j-1)p} & -M_{jp} & - M_{(j+1)p} & \dots  & -M_{pp}\\
    \end{bmatrix}\right)} 
\end{split}
\end{align*}
\begin{align*}
    \begin{split}
    = & \frac{\operatorname{det}\left(\begin{bmatrix} 
    M_{11} & \dots & M_{1(i-1)} & M_{1(i+1)} & \dots  & M_{1p}\\
    \vdots & \ddots & \vdots & \vdots & \ddots & \vdots\\
    M_{(j-1)1} & \dots & M_{(j-1)(i-1)} & M_{(j-1)(i+1)} & \dots  & M_{(j-1)p}\\
    M_{i1} & \dots & M_{i(i-1)} & M_{i(i+1)} & \dots  & M_{ip}\\
    M_{(j+1)1} & \dots & M_{(j+1)(i-1)} & M_{(j+1)(i+1)} & \dots  & M_{(j+1)p}\\
    \vdots & \ddots & \vdots & \vdots & \ddots & \vdots\\
    M_{p1} & \dots & M_{p(i-1)} & M_{p(i+1)} & \dots  & M_{pp}
    \end{bmatrix}\right)}{\operatorname{det}\left(\begin{bmatrix} 
    M_{11} & \dots & M_{1(i-1)} & M_{1(i+1)} & \dots  & M_{1p}\\
    \vdots & \ddots & \vdots & \vdots & \ddots & \vdots\\
    M_{(i-1)1} & \dots & M_{(i-1)(i-1)} & M_{(i-1)(i+1)} & \dots  & M_{(i-1)p}\\
    M_{(i+1)1} & \dots & M_{(i+1)(i-1)} & M_{(i+1)(i+1)} & \dots  & M_{(i+1)p}\\
    \vdots & \ddots & \vdots & \vdots & \ddots & \vdots\\
    M_{p1} & \dots & M_{p(i-1)} & M_{p(i+1)} & \dots  & M_{pp}
    \end{bmatrix}\right)} \\
    & \hspace{15ex} [\text{Taking transpose of both matrices}]
\end{split}
\end{align*}
\begin{align} \label{a_ij_calculation}
    \begin{split}
    = & \, (-1)^{(i-j+1)} \cdot \frac{\operatorname{det}\left(\begin{bmatrix} 
    M_{11} & \dots & M_{1(i-1)} & M_{1(i+1)} & \dots  & M_{1p}\\
    \vdots & \ddots & \vdots & \vdots & \ddots & \vdots\\
    M_{(j-1)1} & \dots & M_{(j-1)(i-1)} & M_{(j-1)(i+1)} & \dots  & M_{(j-1)p}\\
    M_{(j+1)1} & \dots & M_{(j+1)(i-1)} & M_{(j+1)(i+1)} & \dots  & M_{(j+1)p}\\
    \vdots & \ddots & \vdots & \vdots & \ddots & \vdots\\
    M_{p1} & \dots & M_{p(i-1)} & M_{p(i+1)} & \dots  & M_{pp}
    \end{bmatrix}\right)}{\operatorname{det}\left(\begin{bmatrix} 
    M_{11} & \dots & M_{1(i-1)} & M_{1(i+1)} & \dots  & M_{1p}\\
    \vdots & \ddots & \vdots & \vdots & \ddots & \vdots\\
    M_{(i-1)1} & \dots & M_{(i-1)(i-1)} & M_{(i-1)(i+1)} & \dots  & M_{(i-1)p}\\
    M_{(i+1)1} & \dots & M_{(i+1)(i-1)} & M_{(i+1)(i+1)} & \dots  & M_{(i+1)p}\\
    \vdots & \ddots & \vdots & \vdots & \ddots & \vdots\\
    M_{p1} & \dots & M_{p(i-1)} & M_{p(i+1)} & \dots  & M_{pp}
    \end{bmatrix}\right)} \\
    &\quad\quad\quad [\text{Moving $M_{i1},\ldots,M_{ip}$ in the numerator matrix from ($j-1$)th row to $i$th row}]
    \end{split}
\end{align}
So, we can see that the form of $a_{ij}$ obtained in equation \eqref{a_ij_calculation} is exactly same as in equation \eqref{a_ij_new}. So, we can say that our claim is correct and we can get $a_{ij}(i \neq j)$ from the $\textbf{Beta}$ matrix in the above described manner.

\paragraph{\textbf{Note:} \normalfont{If the number of instruments is greater than the number of response variables, i.e. $k > p$, we will select one valid instrument corresponding to each response variable. Consequently, we will create a new matrix called $\textbf{Beta}\_{\text{\textbf{New}}}$ by stacking the columns of the $\textbf{Beta}$ matrix that correspond to these instrument variables. This can always be done because we assume that each response variable is influenced by at least one valid instrument variable (refer to Section (\ref{Identifiable})). The $\textbf{Beta}\_{\text{\textbf{New}}}$ matrix will have dimensions $p \times p$. We will then apply the algorithm described in equation (\ref{a_ij}) to estimate the $\mathbf{A}-$matrix using $\textbf{Beta}\_{\text{\textbf{New}}}$ matrix. The proof for this approach is exactly similar to the one provided earlier.}}

\section{Calculation of $\mathbf{Syx}$ and $\mathbf{Syy}$ from $\mathbf{Sxx}$, $\textbf{Beta}$ \text{and} $\textbf{Sigma}\textbf{Hat}$} \label{S_YX & S_YY}

Suppose we have $\mathbf{Y=(Y_1,\cdots,Y_p)}^{T}$ and $\mathbf{X=(X_1,\cdots,X_k)}^{T}$. $Y_1,\cdots,Y_p$ and $X_1,\cdots,X_k$ are vectors of length $n$, where $n$ is the umber of observations. Let
$\hat{\beta}_{ij}$ i.e. $(i, j)$th entry of $\textbf{Beta}$, denotes the regression coefficient of $\mathbf{Y_i}$ on $\mathbf{X_j}$, where the regression is performed without including an intercept term. So, we have $\hat{\beta}_{ij}$ as follows:
\begin{align*}
    \hat{\beta}_{ij} = \frac{\frac{1}{n}\sum_{t=1}^{n}{Y_{it}X_{jt}}}{\frac{1}{n}\sum_{t=1}^{n}{X_{jt}^2}}
\end{align*}
We have, $\hat{\sigma}_{ij}^2$ i.e. $(i, j)$th entry of $\textbf{Sigma}\textbf{Hat}$ as follows:
\begin{align*}
    \hat{\sigma}_{ij}^2 = \frac{1}{n}\sum_{t=1}^{n}(Y_{it} - \hat{\beta}_{ij}X_{jt})^2
\end{align*}
We also know $\mathbf{Sxx}$, which is a $k\times k$ matrix, whose $(i,j)$th entry is given as:
$$(\mathbf{Sxx})_{(i,j)} = \frac{1}{n}\sum_{t=1}^{n}X_{it}X_{jt}$$
First we want to calculate $\mathbf{Syx}$, which will be a $p\times k$ matrix, whose $(i, j)$ th entry is given as:
$$(\mathbf{Syx})_{(i,j)} = \frac{1}{n}\sum_{t=1}^{n}Y_{it}X_{jt}$$
We have:
\begin{align*}
    & \hat{\beta}_{ij} = (\mathbf{X_j}^{T}\mathbf{X_j})^{-1}\mathbf{X_j}^{T}\mathbf{Y_i}\\
    \implies & \hat{\beta}_{ij} = \left(\frac{1}{n}\sum_{t=1}^{n}X_{jt}X_{jt}\right)^{-1} \times \left(\frac{1}{n}\sum_{t=1}^{n}Y_{it}X_{jt}\right)\\
    \implies & (\mathbf{Syx})_{(i,j)} = \hat{\beta}_{ij} \times (\mathbf{Sxx})_{(j,j)}
\end{align*}
In this way we can calculate the $\mathbf{Syx}$ matrix. Now, we want to calculate $\mathbf{Syy}$ matrix, which will be a $p\times p$ matrix, whose $(i, j)$ th entry is given as:
$$(\mathbf{Syy})_{(i,j)} = \frac{1}{n}\sum_{t=1}^{n}Y_{it}Y_{jt}$$
We will first calculate the diagonal entries of $\mathbf{Syy}$ matrix and then use them to construct the whole matrix. We have,
\begin{align*}
     & \hat{\sigma}_{ij}^2 = \frac{1}{n}\sum_{t=1}^{n}(Y_{it} - \hat{\beta}_{ij}X_{jt})^2,\quad \forall j \in \{1,\cdots,p\}\\
      \implies & \frac{1}{n}\sum_{t=1}^{n} Y_{it}^2 = \hat{\sigma}_{ij}^2 + 2 \times \hat{\beta}_{ij} \times  \left(\frac{1}{n}\sum_{t=1}^{n}Y_{it}X_{jt}\right) - \hat{\beta}_{ij}^2  \left(\frac{1}{n}\sum_{t=1}^{n}X_{jt}^2\right), \quad \forall j \in \{1,\cdots,p\} \\
      \implies & (\mathbf{Syy})_{(i,i)} =  \hat{\sigma}_{ij}^2 +  2 \times \hat{\beta}_{ij} \times (\mathbf{Syx})_{(i,j)} - \hat{\beta}_{ij}^2 \times (\mathbf{Sxx})_{(j,j)}, \quad \forall j \in \{1,\cdots,p\}
\end{align*}
In this way we can calculate the diagonal entries of $\mathbf{Syy}$. We will now try to calculate $\mathbf{Syy}$. We have the following equation:
\begin{align} \label{S_YY}
    & \mathbf{Y_i = AY_i + BX_i + E_i}, \quad i =1, \ldots, n \nonumber \\
    \implies & \mathbf{(I_p - A) Y_i = BX_i + E_i} \nonumber \\
    \implies & \mathbf{Y_i = (I_p - A)^{-1}BX_i + (I_p - A)^{-1} E_i},\quad \text{Var}(\mathbf{E_i}) = \mathbf{\Sigma} =  \begin{bmatrix} 
    \sigma_1^2 & 0 & \dots  & 0\\
    0 & \sigma_2^2 & \dots  & 0\\
    \vdots & \vdots & \ddots & \vdots\\
    0 & 0 & \dots  & \sigma_p^2 
    \end{bmatrix} \nonumber \\
    \implies & \mathbf{E}(\mathbf{YY}^{T}) = (\mathbf{I_p - A})^{-1}\mathbf{B} \times \mathbf{E}(\mathbf{XX}^{T})\times \mathbf{B}^{T}(\mathbf{I_p - A})^{-T} + (\mathbf{I_p - A})^{-1} \times \mathbf{E}(\mathbf{EE}^{T}) \times (\mathbf{I_p - A})^{-T} \nonumber \\
    \implies & \mathbf{Syy} = \textbf{Beta} \times \mathbf{Sxx} \times \textbf{Beta}^{T} + (\mathbf{I_p - A})^{-1} \times \mathbf{\Sigma} \times(\mathbf{I_p - A})^{-T} \quad [\text{By SLLN}]
\end{align}
This is true because earlier we assumed $\textbf{Beta} = (\mathbf{I_p - A})^{-1} \times \mathbf{B}$ in equation (\ref{beta}). We have also shown in Section (\ref{sec:Calculation of A matrix from Beta matrix}) that we can estimate $\mathbf{A}$ matrix from $\textbf{Beta}$ matrix, i.e. we also know $(\mathbf{I_p - A})^{-1}$ term in equation (\ref{S_YY}). So, the only unknown term in equation (\ref{S_YY}) is $\mathbf{\Sigma}$ matrix.\newline
Now we have, 
\begin{align} \label{Y}
    & \mathbf{Y} = (\mathbf{I_p - A})^{-1}\mathbf{BX} + (\mathbf{I_p - A})^{-1}\mathbf{E} \nonumber \\
    \implies & \mathbf{Y = Beta \times X + (\mathbf{I_p - A})^{-1}E}
\end{align}
So, if we consider the first row of the equation (\ref{Y}) we have,
\begin{align*}
    Y_{1i} = \beta_{11}X_{1i} + \cdots +\beta_{1p}X_{pi} + e_{1i}, \quad i = 1,\cdots,n
\end{align*}
We now have, 
\begin{align} \label{R_1sq}
    R_1^2 = 1 - \frac{\sum_{i=1}^ne_{1i}^2}{\sum_{i=1}^nY_{1i}^2}
\end{align}
We know $\sum_{i=1}^nY_{1i}^2$ as this will be the first diagonal entry of $\mathbf{Syy}$. We will now show a way to calculate $R_1^2$ and from that we will try to calculate $\sum_{i = 1}^ne_{1i}^2$.
Let $r_{0i}$ denotes the correlation coefficient between $\mathbf{Y_1}$ and $\mathbf{X_i},i=1,\cdots,k$ and $r_{ij}$ denotes the correlation coefficients between $\mathbf{X_i}$ and $\mathbf{X_j};\,\,i=1,\cdots,k;j=1,\cdots,k;i\neq j$. So, we have the following matrix:
\begin{align*}
   \mathbf{R} =  \begin{bmatrix} 
    1 & r_{01} & r_{02} & \dots  & r_{0p}\\
    r_{01} & 1 & r_{12} & \dots  & r_{1p}\\
    \vdots & \vdots & \vdots & \ddots & \vdots\\
    r_{0p} & r_{p1} & r_{p2} & \dots  & 1 
    \end{bmatrix}
\end{align*}
This matrix can easily be calculated based on $\mathbf{Syx}$ and $\mathbf{Sxx}$. From this $\mathbf{R}$ matrix, $R_1^2$ can be calculated in the following way:
\begin{align} \label{R_1sqnew}
    R_1^2 = 1 - \frac{det(\mathbf{R})}{R_{11}}
\end{align}
where, $R_{11}$ denotes the cofactor of $(1,1)$th element of $\mathbf{R}$. Thus from equation (\ref{R_1sq}) and (\ref{R_1sqnew}) we can calculate $\sum_{i=1}^n e_{1i}^2$. We can say that $\frac{1}{n}\sum_{i=1}^n e_{1i}^2$ will be the MLE of the first diagonal element of $(\mathbf{I_p - A})^{-1} \times \mathbf{\Sigma} \times(\mathbf{I_p - A})^{-T}$. Thus, in this way we can estimate the diagonal entries of $(\mathbf{I_p - A})^{-1} \times \mathbf{\Sigma} \times(\mathbf{I_p - A})^{-T}$. We have shown in Section (\ref{sec:Calculation of A matrix from Beta matrix}) that we can estimate $\mathbf{A}$ from $\textbf{Beta}$ matrix. So, let us assume:
\begin{align*}
    (\mathbf{I_p - A})^{-1} = \begin{bmatrix} 
    M_{11} & M_{12} & \dots  & M_{1p}\\
    M_{21} & M_{22} & \dots  & M_{2p}\\
    \vdots & \vdots & \ddots & \vdots\\
    M_{p1} & M_{p2} & \dots  & M_{pp} 
    \end{bmatrix}
\end{align*}
Thus, equating the diagonal entries of $(\mathbf{I_p - A})^{-1} \times \mathbf{\Sigma} \times(\mathbf{I_p - A})^{-T}$ with their MLEs we have:
\begin{align*}
   & \begin{bmatrix} 
    M_{11}^2 & M_{12}^2 & \dots  & M_{1p}^2\\
    M_{21}^2 & M_{22}^2 & \dots  & M_{2p}^2\\
    \vdots & \vdots & \ddots & \vdots\\
    M_{p1}^2 & M_{p2}^2 & \dots  & M_{pp}^2 
    \end{bmatrix} \times \begin{bmatrix} 
    \sigma_1^2 \\
    \sigma_2^2 \\
    \vdots\\
    \sigma_p^2 \end{bmatrix}
    =
    \begin{bmatrix} 
    \frac{1}{n}\sum_{i=1}^n e_{1i}^2 \\
    \frac{1}{n}\sum_{i=1}^n e_{2i}^2 \\
    \vdots\\
    \frac{1}{n}\sum_{i=1}^n e_{pi}^2 
    \end{bmatrix}\\
    \implies &
    \begin{bmatrix} 
    \sigma_1^2 \\
    \sigma_2^2 \\
    \vdots\\
    \sigma_p^2 \end{bmatrix}
    = \begin{bmatrix} 
    M_{11}^2 & M_{12}^2 & \dots  & M_{1p}^2\\
    M_{21}^2 & M_{22}^2 & \dots  & M_{2p}^2\\
    \vdots & \vdots & \ddots & \vdots\\
    M_{p1}^2 & M_{p2}^2 & \dots  & M_{pp}^2 \end{bmatrix}^{-1} \times 
    \begin{bmatrix} 
    \frac{1}{n}\sum_{i=1}^n e_{1i}^2 \\
    \frac{1}{n}\sum_{i=1}^n e_{2i}^2 \\
    \vdots\\
    \frac{1}{n}\sum_{i=1}^n e_{pi}^2 
    \end{bmatrix}
\end{align*}
As, we have estimated $\mathbf{\Sigma}$ matrix we can now estimate $\mathbf{Syy}$ from equation (\ref{Y}). We have used the idea of Section (\ref{sec:Calculation of A matrix from Beta matrix}) and (\ref{S_YX & S_YY}), to calculate $\mathbf{Syy}$ and $\mathbf{Syx}$, when they are not easily available, from $\mathbf{Sxx}$, $\textbf{Beta}$ and $\textbf{Sigma}\textbf{Hat}$.

\section{Details of the Main Simulation results}
Utilizing the mathematical model $\mathbf{Y_i = AY_i + BX_i + E_i}$, where $\mathbf{B}$ is set as the identity matrix, we conducted a comprehensive assessment of our algorithm's performance. By generating $\mathbf{X_i}$ and $\mathbf{E_i}$ from a normal distribution, we simulated various scenarios to gauge the effectiveness of our approach. Specifically, we calculated True Positive Rate (TPR), False Positive Rate (FPR), False Discovery Rate (FDR), Matthew's Correlation Coefficient (MCC), and Area Under the Curve (AUC) for \textbf{MR.RGM} for both the \textbf{Spike and Slab} prior and the \textbf{Threshold} prior. These metrics were then compared with those obtained through \textbf{OneSampleMR} \cite{OneSampleMR} R package. Our analyses encompassed a range of sample sizes – $10$k, $30$k, and $50$k – as well as different variance explained levels ($1\%, 3\%, 5\% \,\, \text{and} \,\, 10\%$). The study also accounted for network sizes of 5, and 10, network sparsity of $\mathbf{A}$ at $25\%$ and $50\%$, and response variable interactions of either $0.1$ or $-0.1$. A detailed comparison of these factors is presented in table (\ref{tab:Tab 1}), (\ref{tab:Tab 2}), (\ref{tab:Tab 3}) and (\ref{tab:Tab 4}). Notably, our approach consistently yielded equivalent results when applied to both individual level data and summary level data inputs, further validating its versatility and reliability across different data availability scenarios.
\begin{table}[H]
\centering
\tiny 
\caption{Comparison of AUC, TPR, FPR, FDR and MCC between MR.RGM (both with threshold prior and spike and slab prior) and OneSampleMR (for different $\alpha$) for network size $5$ and sparsity $50\%$.}
\label{tab:Tab 1}
\begin{tabular}{|c|c|c|c|c|c|c|c|c|c|c|c|c|c|}
\hline
Sample Size (in k) &  & $10$ & $30$ & $50$ & $10$ & $30$ & $50$ & $10$ & $30$ & $50$ & $10$ & $30$ & $50$  \\
Variance explained (in $\%$) & & $1$ & $1$ & $1$ & $3$ & $3$ & $3$ & $3$ & $5$ & $5$ & $5$ & $10$ & $10$ \\
Network size & & $5$ & $5$ & $5$ & $5$ & $5$ & $5$ & $5$ & $5$ & $5$ & $5$ & $5$ & $5$  \\
Sparsity (in $\%$) & & $50$ & $50$ & $50$ & $50$ & $50$ & $50$ & $50$ & $50$ & $50$ & $50$ & $50$ & $50$  \\
\hline
\hline
 & Threshold prior & $0.67$ & $0.76$ & $0.81$ & $0.82$ & $0.87$ & $0.89$ & $0.87$ & $0.91$ & $0.92$ & $0.998$ & $0.97$ & $0.97$   \\
AUC & Spike $\&$ slab prior & $0.77$ & $0.86$ & $0.92$ & $0.91$ & $0.99$ & $0.98$ & $0.96$ & $0.98$ & $0.99$ & $0.99$ & $0.99$ & $0.99$ \\
 & OneSampleMR & $0.56$ & $0.67$ & $0.74$ & $0.69$ & $0.76$ & $0.79$ & $0.72$ & $0.78$ & $0.79$ & $0.79$ & $0.79$ & $0.78$  \\
 \hline
 \hline
 & TPR & $0.49$ & $0.65$ & $0.75$ & $0.65$ & $0.79$ & $0.83$ & $0.73$ & $0.81$ & $0.85$ & $0.94$ & $0.9$ & $0.93$   \\
Threshold prior & FPR & $0.16$ & $0.19$ & $0.26$ & $0.12$ & $0.08$ & $0.1$ & $0.05$ & $0.06$ & $0.05$ & $0.01$ & $0.02$ & $0.03$ \\
 & FDR & $0.25$ & $0.21$ & $0.24$ & $0.14$ & $0.10$ & $0.12$ & $0.07$ & $0.08$ & $0.07$ & $0.01$ & $0.03$ & $0.03$  \\
 & MCC & $0.35$ & $0.48$ & $0.51$ & $0.57$ & $0.72$ & $0.74$ & $0.72$ & $0.77$ & $0.81$ & $0.94$ & $0.87$ & $0.90$  \\
 \hline
 \hline
  & TPR & $0.5$ & $0.62$ & $0.73$ & $0.63$ & $0.92$ & $0.98$ & $0.8$ & $0.97$ & $0.98$ & $0.97$ & $0.99$ & $0.99$
   \\
Spike $\&$ slab prior & FPR & $0.096$ & $0.05$ & $0.04$ & $0.04$ & $0.009$ & $0.006$ & $0.02$ & $0.001$ & $0.002$ & $0.006$ & $0$ & $0$
 \\
 & FDR & $0.15$ & $0.07$ & $0.05$ & $0.07$ & $0.009$ & $0.006$ & $0.02$ & $0.001$ & $0.002$ & $0.006$ & $0$ & $0$
  \\
 & MCC & $0.44$ & $0.61$ & $0.72$ & $0.62$ & $0.91$ & $0.97$ & $0.8$ & $0.97$ & $0.98$ & $0.96$ & $0.99$ & $0.99$
  \\
 \hline
 \hline
 & TPR & $0.08$ & $0.22$ & $0.39$ & $0.22$ & $0.59$ & $0.76$ & $0.37$ & $0.74$ & $0.80$ & $0.64$ & $0.81$ & $0.81$
\\
OneSampleMR ($\alpha = 0.025$) & FPR & $0.04$ & $0.07$ & $0.09$ & $0.05$ & $0.17$ & $0.20$ & $0.11$ & $0.21$ & $0.23$ & $0.16$ & $0.23$ & $0.24$
 \\
 & FDR & $0.35$ & $0.23$ & $0.18$ & $0.17$ & $0.22$ & $0.20$ & $0.23$ & $0.22$ & $0.22$ & $0.19$ & $0.22$ & $0.23$
  \\
 & MCC & $0.08$ & $0.22$ & $0.35$ & $0.26$ & $0.44$ & $0.57$ & $0.30$ & $0.54$ & $0.58$ & $0.50$ & $0.57$ & $0.57$
  \\
 \hline
 \hline
  & TPR & $0.11$ & $0.33$ & $0.49$ & $0.32$ & $0.66$ & $0.78$ & $0.46$ & $0.77$ & $0.81$ & $0.70$ & $0.81$ & $0.82$
\\
OneSampleMR ($\alpha = 0.05$) & FPR & $0.07$ & $0.11$ & $0.14$ & $0.09$ & $0.21$ & $0.24$ & $0.15$ & $0.25$ & $0.25$ & $0.20$ & $0.27$ & $0.28$
 \\
 & FDR & $0.40$ & $0.24$ & $0.22$ & $0.21$ & $0.24$ & $0.23$ & $0.25$ & $0.24$ & $0.23$ & $0.22$ & $0.24$ & $0.25$
  \\
 & MCC & $0.07$ & $0.26$ & $0.37$ & $0.29$ & $0.46$ & $0.54$ & $0.34$ & $0.52$ & $0.57$ & $0.50$ & $0.55$ & $0.55$
  \\
 \hline
 \hline
  & TPR & $0.20$ & $0.45$ & $0.59$ & $0.43$ & $0.72$ & $0.80$ & $0.58$ & $0.80$ & $0.82$ & $0.77$ & $0.83$ & $0.84$
   \\
OneSampleMR ($\alpha = 0.10$) & FPR & $0.13$ & $0.19$ & $0.21$ & $0.15$ & $0.28$ & $0.29$ & $0.22$ & $0.30$ & $0.30$ & $0.26$ & $0.32$ & $0.34$
 \\
 & FDR & $0.38$ & $0.29$ & $0.26$ & $0.25$ & $0.27$ & $0.26$ & $0.27$ & $0.26$ & $0.26$ & $0.25$ & $0.28$ & $0.28$
  \\
 & MCC &$0.10$ & $0.29$ & $0.39$ & $0.31$ & $0.45$ & $0.52$ & $0.37$ & $0.51$ & $0.53$ & $0.52$ & $0.51$ & $0.51$
  \\
 \hline
 \hline
\end{tabular}
\end{table}

\begin{table}[H]
\centering
\tiny 
\caption{Comparison of AUC, TPR, FPR, FDR and MCC between MR.RGM (both with threshold prior and spike and slab prior) and OneSampleMR (for different $\alpha$) for network size $5$ and sparsity $25\%$.}

\label{tab:Tab 2}
\begin{tabular}{|c|c|c|c|c|c|c|c|c|c|c|c|c|c|}
\hline
Sample Size (in k) &  & $10$ & $30$ & $50$ & $10$ & $30$ & $50$ & $10$ & $30$ & $50$ & $10$ & $30$ & $50$  \\
Variance explained (in $\%$) & & $1$ & $1$ & $1$ & $3$ & $3$ & $3$ & $3$ & $5$ & $5$ & $5$ & $10$ & $10$ \\
Network size & & $5$ & $5$ & $5$ & $5$ & $5$ & $5$ & $5$ & $5$ & $5$ & $5$ & $5$ & $5$  \\
Sparsity (in $\%$) & & $25$ & $25$ & $25$ & $25$ & $25$ & $25$ & $25$ & $25$ & $25$ & $25$ & $25$ & $25$  \\
\hline
\hline
 & Threshold prior & $0.71$ & $0.77$ & $0.76$ & $0.78$ & $0.86$ & $0.86$ & $0.87$ & $0.87$ & $0.90$ & $0.97$ & $0.93$ & $0.94$   \\
AUC & Spike $\&$ slab prior & $0.89$ & $0.94$ & $0.98$ & $0.97$ & $0.99$ & $0.997$ & $0.98$ & $0.99$ & $0.997$ & $0.998$ & $0.998$ & $0.996$
 \\
 & OneSampleMR & $0.53$ & $0.57$ & $0.57$ & $0.57$ & $0.60$ & $0.59$ & $0.58$ & $0.61$ & $0.60$ & $0.59$ & $0.62$ & $0.63$
  \\
 \hline
 \hline
 & TPR & $0.57$ & $0.71$ & $0.74$ & $0.67$ & $0.80$ & $0.83$ & $0.81$ & $0.81$ & $0.85$ & $0.94$ & $0.90$ & $0.90$   \\
Threshold prior & FPR & $0.20$ & $0.22$ & $0.28$ & $0.13$ & $0.14$ & $0.12$ & $0.08$ & $0.11$ & $0.07$ & $0.02$ & $0.03$ & $0.04$ \\
 & FDR &  $0.49$ & $0.45$ & $0.51$ & $0.37$ & $0.34$ & $0.31$ & $0.23$ & $0.26$ & $0.2$ & $0.06$ & $0.1$ & $0.12$   \\
 & MCC & $0.36$ & $0.47$ & $0.44$ & $0.53$ & $0.64$ & $0.67$ & $0.71$ & $0.71$ & $0.79$ & $0.92$ & $0.88$ & $0.85$\\
 \hline
 \hline
  & TPR & $0.7$ & $0.84$ & $0.91$ & $0.86$ & $0.97$ & $0.996$ & $0.93$ & $0.99$ & $0.99$ & $0.988$ & $0.998$ & $0.996$
   \\
Spike $\&$ slab prior & FPR & $0.13$ & $0.07$ & $0.03$ & $0.05$ & $0.018$ & $0.004$ & $0.026$ & $0.008$ & $0.002$ & $0.01$ & $0.001$ & $0.001$
 \\
 & FDR & $0.34$ & $0.19$ & $0.1$ & $0.15$ & $0.05$ & $0.01$ & $0.07$ & $0.02$ & $0.006$ & $0.01$ & $0.002$ & $0.004$
  \\
 & MCC & $0.56$ & $0.76$ & $0.87$ & $0.81$ & $0.95$ & $0.99$ & $0.9$ & $0.98$ & $0.99$ & $0.98$ & $0.98$ & $0.995$
  \\
 \hline
 \hline
 & TPR & $0.07$ & $0.15$ & $0.20$ & $0.14$ & $0.35$ & $0.40$ & $0.22$ & $0.40$ & $0.42$ & $0.35$ & $0.43$ & $0.43$
   \\
OneSampleMR ($\alpha = 0.025$) & FPR & $0.04$ & $0.08$ & $0.12$ & $0.08$ & $0.19$ & $0.21$ & $0.12$ & $0.21$ & $0.22$ & $0.19$ & $0.23$ & $0.22$
 \\
 & FDR & $0.64$ & $0.62$ & $0.65$ & $0.63$ & $0.62$ & $0.61$ & $0.63$ & $0.61$ & $0.61$ & $0.61$ & $0.61$ & $0.61$
 \\
 & MCC & $0.08$ & $0.11$ & $0.10$ & $0.09$ & $0.16$ & $0.19$ & $0.12$ & $0.18$ & $0.19$ & $0.16$ & $0.19$ & $0.20$
  \\
 \hline
 \hline
  & TPR & $0.09$ & $0.20$ & $0.27$ & $0.19$ & $0.39$ & $0.42$ & $0.27$ & $0.42$ & $0.44$ & $0.38$ & $0.45$ & $0.46$
   \\
OneSampleMR ($\alpha = 0.05$) & FPR & $0.07$ & $0.13$ & $0.16$ & $0.13$ & $0.22$ & $0.24$ & $0.16$ & $0.25$ & $0.25$ & $0.22$ & $0.25$ & $0.25$
 \\
 & FDR & $0.71$ & $0.63$ & $0.63$ & $0.68$ & $0.63$ & $0.62$ & $0.65$ & $0.62$ & $0.62$ & $0.63$ & $0.62$ & $0.62$
  \\
 & MCC & $0.04$ & $0.10$ & $0.13$ & $0.07$ & $0.17$ & $0.18$ & $0.11$ & $0.17$ & $0.18$ & $0.15$ & $0.17$ & $0.20$
  \\
 \hline
 \hline
  & TPR & $0.16$ & $0.27$ & $0.36$ & $0.27$ & $0.42$ & $0.46$ & $0.34$ & $0.47$ & $0.47$ & $0.44$ & $0.49$ & $0.50$
  \\
OneSampleMR ($\alpha = 0.10$) & FPR & $0.13$ & $0.21$ & $0.23$ & $0.19$ & $0.28$ & $0.28$ & $0.22$ & $0.30$ & $0.29$ & $0.28$ & $0.30$ & $0.29$
 \\
 & FDR & $0.69$ & $0.68$ & $0.65$ & $0.67$ & $0.66$ & $0.64$ & $0.66$ & $0.64$ & $0.64$ & $0.65$ & $0.64$ & $0.63$
  \\
 & MCC & $0.05$ & $0.07$ & $0.13$ & $0.09$ & $0.13$ & $0.17$ & $0.12$ & $0.16$ & $0.17$ & $0.15$ & $0.17$ & $0.19$
  \\
 \hline
 \hline
\end{tabular}
\end{table}

\begin{table}[H]
\centering
\tiny 
\caption{Comparison of AUC, TPR, FPR, FDR and MCC between MR.RGM (both with threshold prior and spike and slab prior) and OneSampleMR (for different $\alpha$) for network size $10$ and sparsity $50\%$.}
\label{tab:Tab 3}
\begin{tabular}{|c|c|c|c|c|c|c|c|c|c|c|c|c|c|}
\hline
Sample Size (in k) &  & $10$ & $30$ & $50$ & $10$ & $30$ & $50$ & $10$ & $30$ & $50$ & $10$ & $30$ & $50$  \\
Variance explained (in $\%$) & & $1$ & $1$ & $1$ & $3$ & $3$ & $3$ & $3$ & $5$ & $5$ & $5$ & $10$ & $10$ \\
Network size & & $10$ & $10$ & $10$ & $10$ & $10$ & $10$ & $10$ & $10$ & $10$ & $10$ & $10$ & $10$  \\
Sparsity (in $\%$) & & $50$ & $50$ & $50$ & $50$ & $50$ & $50$ & $50$ & $50$ & $50$ & $50$ & $50$ & $50$  \\
\hline
\hline
 & Threshold prior & $0.66$ & $0.78$ & $0.84$ & $0.83$ & $0.93$ & $0.94$ & $0.91$ & $0.96$ & $0.94$ & $0.98$ & $0.98$ & $0.98$  \\
AUC & Spike $\&$ slab prior & $0.69$ & $0.83$ & $0.9$ & $0.87$ & $0.98$ & $0.98$ & $0.95$ & $0.99$ & $0.99$ & $0.99$ & $0.99$ & $0.99$
 \\
 & OneSampleMR & $0.53$ & $0.57$ & $0.59$ & $0.57$ & $0.61$ & $0.61$ & $0.59$ & $0.61$ & $0.62$ & $0.59$ & $0.62$ & $0.61$
  \\
 \hline
 \hline
 & TPR & $0.65$ & $0.84$ & $0.9$ & $0.78$ & $0.95$ & $0.96$ & $0.86$ & $0.98$ & $0.97$ & $0.96$ & $0.98$ & $0.99$   \\
Threshold prior & FPR & $0.45$ & $0.58$ & $0.62$ & $0.21$ & $0.27$ & $0.35$ & $0.14$ & $0.17$ & $0.22$ & $0.03$ & $0.04$ & $0.05$ \\
 & FDR & $0.4$ & $0.4$ & $0.39$ & $0.2$ & $0.21$ & $0.25$ & $0.13$ & $0.14$ & $0.17$ & $0.03$ & $0.04$ & $0.05$  \\
 & MCC & $0.21$ & $0.3$ & $0.36$ & $0.57$ & $0.71$ & $0.65$ & $0.73$ & $0.82$ & $0.77$ & $0.92$ & $0.94$ & $0.93$  \\
 \hline
 \hline
  & TPR & $0.43$ & $0.59$ & $0.7$ & $0.64$ & $0.91$ & $0.96$ & $0.78$ & $0.97$ & $0.98$ & $0.95$ & $0.98$ & $0.98$
   \\
Spike $\&$ slab prior & FPR & $0.13$ & $0.08$ & $0.06$ & $0.06$ & $0.02$ & $0.006$ & $0.03$ & $0.005$ & $0.002$ & $0.01$ & $0.001$ & $0.001$
 \\
 & FDR & $0.23$ & $0.12$ & $0.078$ & $0.08$ & $0.02$ & $0.006$ & $0.04$ & $0.005$ & $0.002$ & $0.01$ & $0.002$ & $0.001$
  \\
 & MCC & $0.34$ & $0.54$ & $0.67$ & $0.61$ & $0.89$ & $0.96$ & $0.77$ & $0.97$ & $0.98$ & $0.94$ & $0.98$ & $0.98$
  \\
 \hline
 \hline
 & TPR & $0.12$ & $0.27$ & $0.38$ & $0.26$ & $0.54$ & $0.62$ & $0.39$ & $0.62$ & $0.62$ & $0.56$ & $0.66$ & $0.67$
  \\
OneSampleMR ($\alpha = 0.05$) & FPR & $0.09$ & $0.19$ & $0.26$ & $0.19$ & $0.35$ & $0.41$ & $0.26$ & $0.41$ & $0.44$ & $0.37$ & $0.45$ & $0.46$
 \\
 & FDR & $0.44$ & $0.41$ & $0.40$ & $0.41$ & $0.39$ & $0.40$ & $0.40$ & $0.39$ & $0.40$ & $0.40$ & $0.40$ & $0.41$
  \\
 & MCC & $0.04$ & $0.10$ & $0.14$ & $0.09$ & $0.20$ & $0.21$ & $0.14$ & $0.22$ & $0.21$ & $0.19$ & $0.22$ & $0.21$
  \\
 \hline
 \hline
  & TPR & $0.20$ & $0.36$ & $0.48$ & $0.36$ & $0.60$ & $0.65$ & $0.47$ & $0.66$ & $0.68$ & $0.61$ & $0.69$ & $0.70$
   \\
OneSampleMR ($\alpha = 0.10$) & FPR & $0.16$ & $0.25$ & $0.34$ & $0.26$ & $0.41$ & $0.46$ & $0.34$ & $0.45$ & $0.49$ & $0.43$ & $0.49$ & $0.52$
 \\
 & FDR & $0.45$ & $0.41$ & $0.41$ & $0.42$ & $0.40$ & $0.41$ & $0.41$ & $0.41$ & $0.42$ & $0.41$ & $0.42$ & $0.42$
  \\
 & MCC & $0.05$ & $0.12$ & $0.15$ & $0.10$ & $0.19$ & $0.20$ & $0.14$ & $0.20$ & $0.20$ & $0.18$ & $0.20$ & $0.20$
 \\
 \hline
 \hline
  & TPR & $0.26$ & $0.44$ & $0.54$ & $0.43$ & $0.64$ & $0.68$ & $0.54$ & $0.69$ & $0.71$ & $0.64$ & $0.71$ & $0.72$
   \\
OneSampleMR ($\alpha = 0.15$) & FPR & $0.22$ & $0.32$ & $0.39$ & $0.33$ & $0.45$ & $0.50$ & $0.39$ & $0.49$ & $0.52$ & $0.47$ & $0.53$ & $0.56$
 \\
 & FDR & $0.46$ & $0.42$ & $0.42$ & $0.43$ & $0.41$ & $0.42$ & $0.42$ & $0.42$ & $0.42$ & $0.42$ & $0.43$ & $0.44$
  \\
 & MCC & $0.05$ & $0.12$ & $0.15$ & $0.11$ & $0.19$ & $0.19$ & $0.14$ & $0.19$ & $0.20$ & $0.18$ & $0.19$ & $0.18$
  \\
 \hline
 \hline
\end{tabular}
\end{table}

\begin{table}[H]
\centering
\tiny 
\caption{Comparison of AUC, TPR, FPR, FDR and MCC between MR.RGM (both with threshold prior and spike and slab prior) and OneSampleMR (for different $\alpha$) for network size $10$ and sparsity $25\%$.}
\label{tab:Tab 4}
\begin{tabular}{|c|c|c|c|c|c|c|c|c|c|c|c|c|c|}
\hline
Sample Size(in k) &  & $10$ & $30$ & $50$ & $10$ & $30$ & $50$ & $10$ & $30$ & $50$ & $10$ & $30$ & $50$  \\
Variance explained(in $\%$) & & $1$ & $1$ & $1$ & $3$ & $3$ & $3$ & $3$ & $5$ & $5$ & $5$ & $10$ & $10$ \\
Network size & & $10$ & $10$ & $10$ & $10$ & $10$ & $10$ & $10$ & $10$ & $10$ & $10$ & $10$ & $10$  \\
Sparsity(in $\%$) & & $25$ & $25$ & $25$ & $25$ & $25$ & $25$ & $25$ & $25$ & $25$ & $25$ & $25$ & $25$  \\
\hline
\hline
 & Threshold prior & $0.74$ & $0.83$ & $0.84$ & $0.87$ & $0.92$ & $0.9$ & $0.93$ & $0.95$ & $0.93$ & $0.98$ & $0.99$ & $0.98$   \\
AUC & Spike $\&$ slab prior & $0.87$ & $0.94$ & $0.97$ & $0.96$ & $0.99$ & $0.99$ & $0.98$ & $0.995$ & $0.99$ & $0.99$ & $0.994$ & $0.995$
 \\
 & OneSampleMR & $0.52$ & $0.49$ & $0.49$ & $0.50$ & $0.49$ & $0.49$ & $0.51$ & $0.50$ & $0.49$ & $0.49$ & $0.49$ & $0.48$
  \\
 \hline
 \hline
 & TPR & $0.7$ & $0.87$ & $0.89$ & $0.84$ & $0.95$ & $0.93$ & $0.91$ & $0.98$ & $0.94$ & $0.98$ & $0.99$ & $0.98$
   \\
Threshold prior & FPR & $0.33$ & $0.42$ & $0.45$ & $0.16$ & $0.23$ & $0.25$ & $0.1$ & $0.12$ & $0.16$ & $0.02$ & $0.04$ & $0.05$ \\
 & FDR & $0.34$ & $0.41$ & $0.4$ & $0.62$ & $0.64$ & $0.61$ & $0.77$ & $0.79$ & $0.7$ & $0.94$ & $0.93$ & $0.89$  \\
 & MCC & $0.56$ & $0.56$ & $0.57$ & $0.35$ & $0.4$ & $0.4$ & $0.23$ & $0.26$ & $0.33$ & $0.07$ & $0.09$ & $0.1$  \\
 \hline
 \hline
  & TPR & $0.69$ & $0.8$ & $0.88$ & $0.83$ & $0.96$ & $0.99$ & $0.9$ & $0.99$ & $0.99$ & $0.98$ & $0.99$ & $0.99$   \\
Spike $\&$ slab prior & FPR & $0.11$ & $0.07$ & $0.04$ & $0.05$ & $0.013$ & $0.006$ & $0.03$ & $0.004$ & $0.002$ & $0.009$ & $0.001$ & $0.001$
 \\
 & FDR & $0.31$ & $0.19$ & $0.11$ & $0.14$ & $0.04$ & $0.017$ & $0.08$ & $0.01$ & $0.008$ & $0.02$ & $0.004$ & $0.004$
  \\
 & MCC & $0.58$ & $0.73$ & $0.84$ & $0.79$ & $0.95$ & $0.98$ & $0.88$ & $0.99$ & $0.99$ & $0.97$ & $0.99$ & $0.99$
  \\
 \hline
 \hline
 & TPR & $0.07$ & $0.13$ & $0.19$ & $0.14$ & $0.25$ & $0.29$ & $0.20$ & $0.29$ & $0.30$ & $0.26$ & $0.30$ & $0.31$
   \\
OneSampleMR ($\alpha = 0.05$) & FPR & $0.08$ & $0.14$ & $0.20$ & $0.14$ & $0.26$ & $0.29$ & $0.19$ & $0.29$ & $0.31$ & $0.26$ & $0.31$ & $0.33$
 \\
 & FDR & $0.75$ & $0.77$ & $0.75$ & $0.75$ & $0.75$ & $0.75$ & $0.74$ & $0.75$ & $0.75$ & $0.75$ & $0.75$ & $0.75$
  \\
 & MCC & $0.01$ & $0.02$ & $0.01$ & $0.00$ & $0.01$ & $0.01$ & $0.00$ & $0.00$ & $0.01$ & $0.00$ & $0.00$ & $0.01$
  \\
 \hline
 \hline
  & TPR & $0.14$ & $0.20$ & $0.26$ & $0.21$ & $0.30$ & $0.33$ & $0.27$ & $0.33$ & $0.34$ & $0.32$ & $0.34$ & $0.35$
   \\
OneSampleMR ($\alpha = 0.10$) & FPR & $0.13$ & $0.22$ & $0.27$ & $0.21$ & $0.31$ & $0.34$ & $0.26$ & $0.34$ & $0.36$ & $0.32$ & $0.35$ & $0.37$
 \\
 & FDR & $0.73$ & $0.76$ & $0.75$ & $0.75$ & $0.75$ & $0.75$ & $0.74$ & $0.75$ & $0.75$ & $0.75$ & $0.75$ & $0.76$
  \\
 & MCC & $0.01$ & $0.02$ & $0.00$ & $0.00$ & $0.01$ & $0.01$ & $0.01$ & $0.00$ & $0.02$ & $0.00$ & $0.02$ & $0.03$
  \\
 \hline
 \hline
  & TPR & $0.21$ & $0.26$ & $0.32$ & $0.27$ & $0.35$ & $0.37$ & $0.32$ & $0.37$ & $0.38$ & $0.36$ & $0.38$ & $0.39$
   \\
OneSampleMR ($\alpha = 0.15$) & FPR & $0.19$ & $0.28$ & $0.32$ & $0.27$ & $0.36$ & $0.39$ & $0.31$ & $0.38$ & $0.40$ & $0.36$ & $0.40$ & $0.42$
 \\
 & FDR & $0.72$ & $0.76$ & $0.75$ & $0.74$ & $0.75$ & $0.75$ & $0.73$ & $0.75$ & $0.75$ & $0.75$ & $0.75$ & $0.76$
  \\
 & MCC & $0.02$ & $0.02$ & $0.00$ & $0.00$ & $0.00$ & $0.01$ & $0.02$ & $0.01$ & $0.02$ & $0.00$ & $0.02$ & $0.03$
  \\
 \hline
 \hline
\end{tabular}
\end{table}


\section{Selection of Instrumental Variables and Real Data Application}

\subsection{Selection of Instrumental Variables in Real Gene Expression Data}

In MR studies, selecting appropriate instrumental variables (IVs) is crucial for obtaining reliable estimates. For researchers applying the \textbf{MR.RGM} method to real gene expression data, the choice of IVs significantly impacts the validity and precision of causal estimates. Below are key recommendations for selecting IVs in the context of gene expression studies:

\begin{enumerate}
    \item \textbf{cis-eQTL Selection:} In gene expression studies, cis-eQTLs (expression quantitative trait loci) are SNPs located within a short distance from the gene of interest (typically within 1 Mb). These SNPs influence the gene's expression levels and are well-suited as natural instruments. By focusing on cis-eQTLs, researchers ensure that the IVs are biologically relevant and specific to the gene under study. This approach leverages the natural variation in gene expression to infer causal effects.

    \item \textbf{Top K SNPs as IVs:} It is advisable to select the top K cis-eQTLs (usually 5 or 10) based on their association strength with gene expression levels. This can be achieved by ranking SNPs according to their p-values or effect sizes from the cis-eQTL analysis. Selecting a manageable number of top SNPs helps in reducing dimensionality while retaining the most impactful IVs, thus simplifying the model without sacrificing essential information. This method is illustrated in the later real data application section.

    \item \textbf{Model Identifiability:} Ensuring that each gene has at least one unique IV—one not shared with other genes—is crucial for model identifiability. 

    \item \textbf{Effect of Instrument Selection:} The choice of strong and valid IVs significantly impacts the accuracy and precision of causal estimates. Weak instruments can lead to biased or imprecise results. Therefore, selecting a reasonable number (e.g., $5$ or $10$) of top SNPs or top PCs is recommended. Empirical evidence from simulations shows that choosing a limited number of top cis-eQTLs per gene yields reliable and robust results.
\end{enumerate}

The implementation of these IV selection strategies in real data, along with a practical example, is presented in the following section.

\subsection{Real Data Application}

We applied \textbf{MR.RGM} to a real gene expression dataset from the GTEx \cite{gtex2020gtex} Analysis V$7$ database, which offers a wealth of information on human gene expression across various tissues. Specifically, we analyzed the muscle skeletal tissues from $332$ donors whose gene expression and SNPs are both available.


For illustration, we selected the following \(15\) genes: 
\texttt{"TWIST1", "FGFBP3", "ITM2C", "FGF12", "ASPSCR1", "TMEM230", "NUDT4", "VPS25", "PAN3", "RSRC1", "PTPN21", "CCT6A", "C1orf56", "PPIL2", and "DUSP18"}.
Through cis-eQTL analysis, we extracted the top $15$ SNPs for each gene. Some SNPs were overlapped, leading to a total of $31$ unique variants.



\begin{figure}[H]
    \centering
    \includegraphics[width=\textwidth]{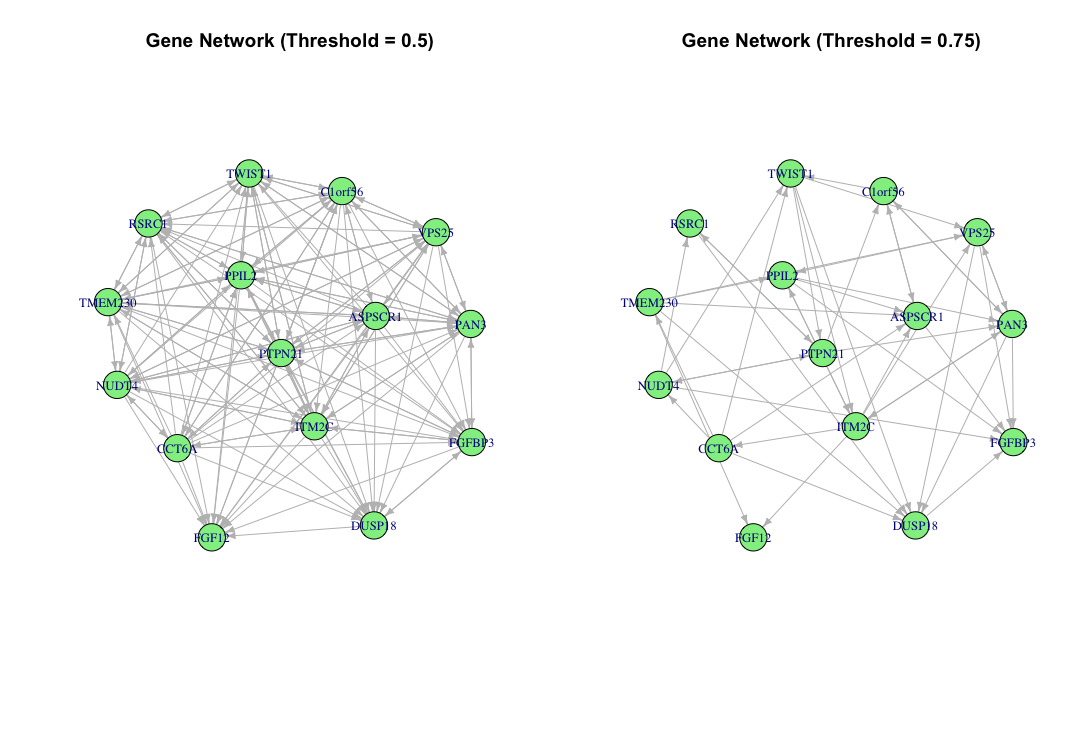}
    \caption{Gene networks constructed using the \textbf{MR.RGM} method with varying thresholds. The left plot shows the network with a threshold of $0.5$, while the right plot depicts the network with a threshold of $0.75$.}
    \label{fig:gene_networks}
\end{figure}

We ran \textbf{MR.RGM} with \textbf{nIter} = $50,000$, \textbf{nBurnin} = $10,000$, and \textbf{Thin} = $10$. The 
runtime was $32.329$ seconds. 
Based on the estimated \textbf{GammaEst}, representing the edge inclusion probabilities, we applied thresholds of $0.5$ and $0.75$ to obtain the gene networks (Figure (\ref{fig:gene_networks})). This approach demonstrates how to construct the input data and apply the \textbf{RGM} methodology effectively.

\color{black}

\newpage

\bibliographystyle{unsrt}
\bibliography{mybib}

\end{document}